\documentclass[11pt]{article}
\usepackage{amsmath,amsthm,amssymb}
\usepackage[a4paper,margin=2.5cm]{geometry}
\usepackage{authblk}
\usepackage{cite}
\usepackage{array}
\usepackage{stackrel}
\usepackage{tikz}
\usepackage{bbold}
\usetikzlibrary{calc,decorations.pathmorphing,decorations.markings,decorations.pathreplacing,patterns,shapes,arrows}
\usetikzlibrary{calc,decorations.pathmorphing,decorations.markings,
  decorations.pathreplacing,patterns,shapes}

\tikzset{boper/.style={rectangle,fill,inner sep=2pt,black}}
\tikzset{bblob/.style={circle,fill,inner sep=1pt,black}}
\tikzset{blob/.style={circle,draw,fill=white,outer sep=0mm,inner sep=0.3mm}}

\tikzset{mblob/.style={circle,fill=white,draw=black,inner sep=1.5pt}}
\tikzset{diam/.style={diamond,fill,inner sep=1.5pt,black}}
\tikzset{pblob/.style={circle,fill=white,draw=red,inner sep=1.5pt}}

\tikzset{
  aline/.style={
    decorate,
    decoration={
      meta-amplitude=#1,
      meta-segment length=0.15cm,
},
    postaction={decorate,ultra thick,decoration={markings,mark = at position #1 with {\arrow{>}}}}        
  },
  aline/.default=0.9
}

\tikzset{
  bline/.style={blue,
    decorate,
    decoration={
      meta-amplitude=#1,
      meta-segment length=0.3cm,
},
    postaction={decorate,ultra thick,decoration={markings,mark = at position #1 with {\arrow{>}}}}        
  },
  bline/.default=0.9
}

\tikzset{edge/.style={very thick,draw=blue}}%
\tikzset{contour/.style={brown,dashed,postaction={decorate,decoration={markings,mark = at position #1 with {\arrow{>}}}}}}
\tikzset{contour/.default=0.5}

\def\loos{0.35}
\def\slt{0.2}
\pgfmathsetmacro{\ae}{atan(\slt)}
\pgfmathsetmacro{\aw}{\ae+180}
\pgfmathsetmacro{\an}{90-\ae}
\pgfmathsetmacro{\as}{\an+180}
\pgfmathsetmacro{\sltb}{sqrt(1-\slt*\slt)}
\pgfmathsetmacro{\lcrot}{45-atan(\slt/\sltb)*0.5}
\tikzset{distort/.style={cm={1,0,-\slt,\sltb,(0,0)}}}
\def\goI#1(#2,#3){
\pgfextra{
\pgfmathparse{#1+180}\global\let\oldangle=\currentangle\global\let\newangle=\pgfmathresult\global\let\currentangle=#1
\pgfmathparse{\oldx+#2}\global\let\newx=\pgfmathresult\xdef\oldx{#2}
\pgfmathparse{\oldy+#3}\global\let\newy=\pgfmathresult\xdef\oldy{#3}
}
.. controls ++(\oldangle:\loos) and ++(\newangle:\loos) .. ++(\newx,\newy)
}
\def\go#1{\expandafter\goI#1}
\def\startI#1(#2,#3){
\pgfextra{\global\let\currentangle=#1\xdef\oldx{#2}\xdef\oldy{#3}}}
\def\start#1{\expandafter\startI#1}
\def\north{\an(0,0.5)}
\def\south{\as(0,-0.5)}
\def\east{\ae(0.5,0)}
\def\west{\aw(-0.5,0)}

\tikzset{bgplaq/.style={fill=lightgray!20!white}}
\tikzset{dgreen/.style={green!50!black,thick}}

\def\plaq(#1,#2){
\begin{scope}[shift={(#1,#2)}]
\draw[dotted] (-0.5,-0.5) rectangle ++(1,1); 
\end{scope}
}

\def\plaqz(#1,#2){
\begin{scope}[shift={(#1,#2)}]
\draw[bgplaq,dotted] (-0.5,-0.5) rectangle ++(1,1); 
\end{scope}
}

\def\plaqa(#1,#2){
\begin{scope}[shift={(#1,#2)}]
\draw[dotted,bgplaq] (-0.5,-0.5) rectangle ++(1,1);
\draw[edge] (0,-0.5) \start\north\go\east;
\draw[edge] (0,0.5) \start\south\go\west;
\end{scope}
}
\def\plaqb(#1,#2){
\begin{scope}[shift={(#1,#2)}]
\draw[dotted,bgplaq] (-0.5,-0.5) rectangle ++(1,1);
\draw[edge] (0,0.5) \start\south\go\east;
\draw[edge] (0,-0.5) \start\north\go\west;
\end{scope}
}

\def\bplaqe(#1,#2){
\begin{scope}[shift={(#1,#2)}]
\draw[dotted] (-0.5,0.5) -- (0.5,0.5) -- (-0.5,-0.5) -- cycle;
\end{scope}
}
\def\bplaqw(#1,#2){
\begin{scope}[shift={(#1,#2)}]
\draw[dotted] (0.5,-0.5) -- (0.5,0.5) -- (-0.5,-0.5) -- cycle;
\end{scope}
}
\def\bplaqn(#1,#2){
\begin{scope}[shift={(#1,#2)}]
\draw[dotted] (0.5,-0.5) -- (-0.5,0.5) -- (-0.5,-0.5) -- cycle;
\end{scope}
}
\def\bplaqs(#1,#2){
\begin{scope}[shift={(#1,#2)}]
\draw[dotted] (0.5,-0.5) -- (-0.5,0.5) -- (0.5,0.5) -- cycle;
\end{scope}
}
\def\bplaqez(#1,#2){
\begin{scope}[shift={(#1,#2)}]
\draw[dotted,bgplaq] (-0.5,0.5) -- (0.5,0.5) -- (-0.5,-0.5) -- cycle;
\end{scope}
}
\def\bplaqwz(#1,#2){
\begin{scope}[shift={(#1,#2)}]
\draw[dotted,bgplaq] (0.5,-0.5) -- (0.5,0.5) -- (-0.5,-0.5) -- cycle;
\end{scope}
}
\def\bplaqnz(#1,#2){
\begin{scope}[shift={(#1,#2)}]
\draw[dotted,bgplaq] (0.5,-0.5) -- (-0.5,0.5) -- (-0.5,-0.5) -- cycle;
\end{scope}
}
\def\bplaqsz(#1,#2){
\begin{scope}[shift={(#1,#2)}]
\draw[dotted,bgplaq] (0.5,-0.5) -- (-0.5,0.5) -- (0.5,0.5) -- cycle;
\end{scope}
}
\def\bplaqea(#1,#2){
\begin{scope}[shift={(#1,#2)}]
\draw[dotted,bgplaq] (-0.5,0.5) -- (0.5,0.5) -- (-0.5,-0.5) -- cycle;
\draw[edge] (0,0.5) \start\south\go\west;
\end{scope}
}
\def\bplaqeb(#1,#2){
\begin{scope}[shift={(#1,#2)}]
\draw[dotted,bgplaq] (-0.5,0.5) -- (0.5,0.5) -- (-0.5,-0.5) -- cycle;
\draw[edge] (0,0.5) \start\south\go\west
node[blob] {};
\end{scope}
}
\def\bplaqwa(#1,#2){
\begin{scope}[shift={(#1,#2)}]
\draw[dotted,bgplaq] (0.5,-0.5) -- (0.5,0.5) -- (-0.5,-0.5) -- cycle;
\draw[edge] (0,-0.5) \start\north\go\east;
\end{scope}
}
\def\bplaqwb(#1,#2){
\begin{scope}[shift={(#1,#2)}]
\draw[dotted,bgplaq] (0.5,-0.5) -- (0.5,0.5) -- (-0.5,-0.5) -- cycle;
\draw[edge] (0,-0.5) \start\north\go\east
node[blob] {};
\end{scope}
}
\def\bplaqna(#1,#2){
\begin{scope}[shift={(#1,#2)}]
\draw[dotted,bgplaq] (0.5,-0.5) -- (-0.5,0.5) -- (-0.5,-0.5) -- cycle;
\draw[edge] (0,-0.5) \start\north\go\west;
\end{scope}
}
\def\bplaqnb(#1,#2){
\begin{scope}[shift={(#1,#2)}]
\draw[dotted,bgplaq] (0.5,-0.5) -- (-0.5,0.5) -- (-0.5,-0.5) -- cycle;
\draw[edge] (0,-0.5) \start\north\go\west 
node[blob] {};
\end{scope}
}
\def\bplaqsa(#1,#2){
\begin{scope}[shift={(#1,#2)}]
\draw[dotted,bgplaq] (0.5,-0.5) -- (-0.5,0.5) -- (0.5,0.5) -- cycle;
\draw[edge] (0,0.5) \start\south\go\east;
\end{scope}
}
\def\bplaqsb(#1,#2){
\begin{scope}[shift={(#1,#2)}]
\draw[dotted,bgplaq] (0.5,-0.5) -- (-0.5,0.5) -- (0.5,0.5) -- cycle;
\draw[edge] (0,0.5) \start\south\go\east
node[blob] {}; 
\end{scope}
}

\def\plaqff(#1,#2){
\begin{scope}[shift={(#1,#2)}]
\draw[dotted,bgplaq] (-0.5,-0.5) rectangle ++(1,1);
\draw[edge] (0,-0.5) \start\north\go\east;
\end{scope}
}
\def\plaqf(#1,#2){
\begin{scope}[shift={(#1,#2)}]
\draw[dotted,bgplaq] (-0.5,-0.5) rectangle ++(1,1);
\draw[edge] (0,0.5) \start\south\go\west;
\end{scope}
}
\def\plaqd(#1,#2){
\begin{scope}[shift={(#1,#2)}]
\draw[dotted,bgplaq] (-0.5,-0.5) rectangle ++(1,1);
\draw[edge] (0,-0.5) \start\north\go\west;
\end{scope}
}
\def\plaqdd(#1,#2){
\begin{scope}[shift={(#1,#2)}]
\draw[dotted,bgplaq] (-0.5,-0.5) rectangle ++(1,1);
\draw[edge] (0,0.5) \start\south\go\east;
\end{scope}
}
\def\plaqc(#1,#2){
\begin{scope}[shift={(#1,#2)}]
\draw[dotted,bgplaq] (-0.5,-0.5) rectangle ++(1,1);
\draw[edge] (0,-0.5) \start\north\go\north;
\end{scope}
}
\def\plaqcc(#1,#2){
\begin{scope}[shift={(#1,#2)}]
\draw[dotted,bgplaq] (-0.5,-0.5) rectangle ++(1,1);
\draw[edge] (-0.5,0) \start\east\go\east;
\end{scope}
}
\def\plaqg(#1,#2){
\begin{scope}[shift={(#1,#2)}]
\draw[dotted,bgplaq] (-0.5,-0.5) rectangle ++(1,1);
\end{scope}
}

\def\cross(#1,#2){
\begin{scope}[shift={(#1,#2)}]
\draw[thick,blue] (-0.1,-0.1) -- (0.1,0.1); 
\draw[thick,blue] (-0.1,0.1) -- (0.1,-0.1); 
\end{scope}
}

\def\gcross(#1,#2){
\begin{scope}[shift={(#1,#2)}]
\draw[thick,green!50!black] (-0.1,-0.1) -- (0.1,0.1); 
\draw[thick,green!50!black] (-0.1,0.1) -- (0.1,-0.1); 
\end{scope}
}

\def\scross(#1,#2){
\begin{scope}[shift={(#1,#2)}]
\draw[thick,blue] (-0.07,-0.07) -- (0.07,0.07); 
\draw[thick,blue] (-0.07,0.07) -- (0.07,-0.07); 
\end{scope}
}

\def\triplaq(#1,#2,#3,#4,#5){
\begin{scope}[shift={(#1,#2)}]
\draw[dgreen] (1.6,-1) -- (0,0) -- (1.6,1);
\draw(0,0) node[bblob] {};
\draw(1.6,1) node[bblob] {};
\draw(1.6,-1) node[bblob] {};
\draw(0,0) node[left] {${#3}$};
\draw(1.6,1) node[right] {${#4}$};
\draw(1.6,-1) node[right] {${#5}$};
\draw[aline=0.9] (0.8,0.5) -- ( (0.8,-0.5);
\cross(0.8,0);
\draw[blue,wavy] (1.6,0) -- (0.8,0);
\end{scope}
}

\def\recpplaq(#1,#2,#3,#4,#5,#6){
\begin{scope}[shift={(#1,#2)}]
\draw[dgreen] (-0.5,-0.5) -- (0.5,-0.5);
\draw[dgreen] (-0.5,0.5) -- (0.5,0.5);
\draw (-0.5,-0.5) node[left] {${#6}$};
\draw (0.5,-0.5)  node[right] {${#5}$};
\draw (-0.5,0.5) node[left] {${#3}$};
\draw  (0.5,0.5) node[right] {${#4}$};
\draw (-0.5,-0.5) node[bblob] {};
\draw (0.5,-0.5)  node[bblob] {};
\draw (-0.5,0.5) node[bblob] {};
\draw  (0.5,0.5) node[bblob] {};
\draw[aline=0.9] (0,0.5) -- ( (0,-0.5);
\draw[blue,wavy=0.2] (-0.5,0) -- (0.5,0);
\end{scope}
}

\def\recmplaq(#1,#2,#3,#4,#5,#6){
\begin{scope}[shift={(#1,#2)}]
\draw[dgreen] (-0.5,-0.5) -- (0.5,-0.5);
\draw[dgreen] (-0.5,0.5) -- (0.5,0.5);
\draw (-0.5,-0.5) node[left] {${#6}$};
\draw (0.5,-0.5)  node[right] {${#5}$};
\draw (-0.5,0.5) node[left] {${#3}$};
\draw  (0.5,0.5) node[right] {${#4}$};
\draw (-0.5,-0.5) node[bblob] {};
\draw (0.5,-0.5)  node[bblob] {};
\draw (-0.5,0.5) node[bblob] {};
\draw  (0.5,0.5) node[bblob] {};
\draw[aline=0.9] (0,0.5) -- ( (0,-0.5);
\draw[blue,wavy=0.2] (0.5,0) -- (-0.5,0);
\end{scope}
}

\def\recpdashplaq(#1,#2,#3,#4,#5,#6){
\begin{scope}[shift={(#1,#2)}]
\draw[dgreen,dashed] (-0.5,-0.5) -- (0.5,-0.5);
\draw[dgreen] (-0.5,0.5) -- (0.5,0.5);
\draw (-0.5,-0.5) node[left] {${#6}$};
\draw (0.5,-0.5)  node[right] {${#5}$};
\draw (-0.5,0.5) node[left] {${#3}$};
\draw  (0.5,0.5) node[right] {${#4}$};
\draw (-0.5,-0.5) node[bblob] {};
\draw (0.5,-0.5)  node[bblob] {};
\draw (-0.5,0.5) node[bblob] {};
\draw  (0.5,0.5) node[bblob] {};
\draw[aline=0.9] (0,0.5) -- ( (0,-0.5);
\draw[blue,wavy=0.2] (-0.5,0) -- (0.5,0);
\end{scope}
}

\def\recmdashplaq(#1,#2,#3,#4,#5,#6){
\begin{scope}[shift={(#1,#2)}]
\draw[dgreen,dashed] (-0.5,-0.5) -- (0.5,-0.5);
\draw[dgreen] (-0.5,0.5) -- (0.5,0.5);
\draw (-0.5,-0.5) node[left] {${#6}$};
\draw (0.5,-0.5)  node[right] {${#5}$};
\draw (-0.5,0.5) node[left] {${#3}$};
\draw  (0.5,0.5) node[right] {${#4}$};
\draw (-0.5,-0.5) node[bblob] {};
\draw (0.5,-0.5)  node[bblob] {};
\draw (-0.5,0.5) node[bblob] {};
\draw  (0.5,0.5) node[bblob] {};
\draw[aline=0.9] (0,0.5) -- ( (0,-0.5);
\draw[blue,wavy=0.2] (0.5,0) -- (-0.5,0);
\end{scope}
}

\def\sosplaq(#1,#2){
\begin{scope}[shift={(#1,#2)}]
\draw[dgreen] (-0.5,-0.5) rectangle ++(1,1);
\draw(0.5,0.5) node[bblob] {};
\draw(0.5,-0.5) node[bblob] {};
\draw(-0.5,0.5) node[bblob] {};
\draw(-0.5,-0.5) node[bblob] {};
\draw[dgreen] (-0.5,0.2) -- (-0.3,0.5);
\end{scope}
}

\def\udashsosplaq(#1,#2){
\begin{scope}[shift={(#1,#2)}]
\draw[dgreen,dashed] (-0.5,0.5) -- (0.5,0.5) -- (0.5,-0.5);
\draw[dgreen] (-0.5,0.5)-- (-0.5,-0.5) -- (0.5,-0.5);
\draw(0.5,0.5) node[bblob] {};
\draw(0.5,-0.5) node[bblob] {};
\draw(-0.5,0.5) node[bblob] {};
\draw(-0.5,-0.5) node[bblob] {};
\draw[dgreen] (-0.5,0.2) -- (-0.3,0.5);
\end{scope}
}

\def\lsosplaq(#1,#2){
\begin{scope}[shift={(#1,#2)}]
\draw[dgreen] (-0.5,-0.5) rectangle (0.5,0.2) ++(1,1);
\draw(0.5,0.2) node[bblob] {};
\draw(0.5,-0.5) node[bblob] {};
\draw(-0.5,0.2) node[bblob] {};
\draw(-0.5,-0.5) node[bblob] {};
\draw[dgreen] (-0.5,-0.1) -- (-0.3,0.2);
\end{scope}
}

\def\usosplaq(#1,#2){
\begin{scope}[shift={(#1,#2)}]
\draw[dgreen] (-0.5,-0.2) rectangle (0.5,0.5) ++(1,1);
\draw(0.5,0.5) node[bblob] {};
\draw(0.5,-0.2) node[bblob] {};
\draw(-0.5,0.5) node[bblob] {};
\draw(-0.5,-0.2) node[bblob] {};
\draw[dgreen] (-0.5,0.2) -- (-0.3,0.5);
\end{scope}
}
\def\lslantsosplaq(#1,#2){
\begin{scope}[shift={(#1,#2)}]
\draw[dgreen] (-0.5,-0.5) -- (0.5,-0.5) -- (0.5,0.2) -- (-0.5,0.5) -- (-0.5,-0.5) ++(1,1);
\draw(0.5,0.2) node[bblob] {};
\draw(0.5,-0.5) node[bblob] {};
\draw(-0.5,0.5) node[bblob] {};
\draw(-0.5,-0.5) node[bblob] {};
\draw[dgreen] (-0.5,0.2) -- (-0.3,0.4);
\end{scope}
}

\def\uslantsosplaq(#1,#2){
\begin{scope}[shift={(#1,#2)}]
\draw[dgreen] (-0.5,-0.5) -- (0.5,-0.2) -- (0.5,0.5) -- (-0.5,0.5) -- (-0.5,-0.5) ++(1,1);
\draw(0.5,0.5) node[bblob] {};
\draw(0.5,-0.2) node[bblob] {};
\draw(-0.5,0.5) node[bblob] {};
\draw(-0.5,-0.5) node[bblob] {};
\draw[dgreen] (-0.5,0.2) -- (-0.3,0.5);
\end{scope}
}

\def\luslantsosplaq(#1,#2){
\begin{scope}[shift={(#1,#2)}]
\draw[dgreen] (-0.5,-0.2) -- (0.5,-0.5) -- (0.5,0.5) -- (-0.5,0.5) -- (-0.5,-0.2) ++(1,1);
\draw(0.5,0.5) node[bblob] {};
\draw(0.5,-0.5) node[bblob] {};
\draw(-0.5,0.5) node[bblob] {};
\draw(-0.5,-0.2) node[bblob] {};
\draw[dgreen] (-0.5,0.2) -- (-0.3,0.5);
\end{scope}
}

\def\rcasosplaq(#1,#2){
\begin{scope}[shift={(#1,#2)}]
\draw[dgreen] (-0.75,-0.5) -- (0.5,-0.5) -- (0.5,0.5) -- (-0.5,0.5) -- (-0.75,-0.5) ++(1,1);
\draw(0.5,0.5) node[bblob] {};
\draw(0.5,-0.5) node[bblob] {};
\draw(-0.5,0.5) node[bblob] {};
\draw(-0.75,-0.5) node[bblob] {};
\draw[dgreen] (-0.6,0.15) -- (-0.3,0.5);
\end{scope}
}

\def\rcbsosplaq(#1,#2){
\begin{scope}[shift={(#1,#2)}]
\draw[dgreen] (-0.5,-0.5) -- (0.75,-0.5) -- (0.5,0.5) -- (-0.5,0.5) -- (-0.5,-0.5) ++(1,1);
\draw(0.5,0.5) node[bblob] {};
\draw(0.75,-0.5) node[bblob] {};
\draw(-0.5,0.5) node[bblob] {};
\draw(-0.5,-0.5) node[bblob] {};
\draw[dgreen] (-0.5,0.2) -- (-0.3,0.5);
\end{scope}
}

\def\rwidelsosplaq(#1,#2){
\begin{scope}[shift={(#1,#2)}]
\draw[dgreen] (-0.5,-0.2) rectangle (1,0.5) ++(1,1);
\draw(1,0.5) node[bblob] {};
\draw(1,-0.2) node[bblob] {};
\draw(-0.5,0.5) node[bblob] {};
\draw(-0.5,-0.2) node[bblob] {};
\draw[dgreen] (-0.5,0.2) -- (-0.3,0.5);
\end{scope}
}

\def\rwidesosplaq(#1,#2){
\begin{scope}[shift={(#1,#2)}]
\draw[dgreen] (-0.5,-0.5) rectangle (1,0.5) ++(1,1);
\draw(1,0.5) node[bblob] {};
\draw(1,-0.5) node[bblob] {};
\draw(-0.5,0.5) node[bblob] {};
\draw(-0.5,-0.5) node[bblob] {};
\draw[dgreen] (-0.5,0.2) -- (-0.3,0.5);
\end{scope}
}

\def\tallsosplaq(#1,#2){
\begin{scope}[shift={(#1,#2)}]
\draw[dgreen] (-0.5,-0.5) rectangle (0.5,0.8) ++(1,1);
\draw(0.5,0.8) node[bblob] {};
\draw(0.5,-0.5) node[bblob] {};
\draw(-0.5,0.8) node[bblob] {};
\draw(-0.5,-0.5) node[bblob] {};
\draw[dgreen] (-0.5,0.5) -- (-0.3,0.8);
\end{scope}
}

\pgfdeclaremetadecoration{prewavy}{zigzag}{
  \state{zigzag}[width=\pgfmetadecorationsegmentamplitude*\pgfmetadecoratedpathlength - 0.5*\pgfmetadecorationsegmentlength ,
                  next state=straight] {
        \decoration{zigzag}
      }
    \state{straight}[width=\pgfmetadecorationsegmentlength,
                   next state=final] {
        \decoration{curveto}
    }
    \state{final}{
        \decoration{zigzag}
        \beforedecoration{\pgfpathmoveto{\pgfpointmetadecoratedpathfirst}}
    }
}

\tikzset{
  wavy/.style={
    decorate,
    decoration={
      prewavy,
      meta-amplitude=#1,
      meta-segment length=0.3cm,
      amplitude=1.5pt, 
      segment length=6pt 
},
    postaction={decorate,ultra thick,decoration={markings,mark = at position #1 with {\arrow{>}}}}        
  },
  wavy/.default=0.5
}
\tikzset{oper/.style={rectangle,fill,inner sep=2.5pt}}
\tikzset{arr/.style={postaction={decorate,thick,decoration={markings,mark = at position #1 with {\arrow{>}}}}}}

\tikzset{
  abline/.style={blue,dashed,
    decorate,
    decoration={
      meta-amplitude=#1,
      meta-segment length=0.3cm,
},
    postaction={decorate,ultra thick,decoration={markings,mark = at position #1 with {\arrow{>}}}}        
  },
  abline/.default=0.5
}

\tikzset{
  aline/.style={
    decorate,
    decoration={
      meta-amplitude=#1,
      meta-segment length=0.3cm,
},
    postaction={decorate,ultra thick,decoration={markings,mark = at position #1 with {\arrow{>}}}}        
  },
  aline/.default=0.5
}

\tikzset{
  agline/.style={green!50!black,thick,
    decorate,
    decoration={
      meta-amplitude=#1,
      meta-segment length=0.3cm,
},
    postaction={decorate,ultra thick,decoration={markings,mark = at position #1 with {\arrow{>}}}}        
  },
  agline/.default=0.5
}

\tikzset{
  gline/.style={color=green!50!black,thick}}

\definecolor{fondo}{rgb}{0.898,0.996,0.898}

\makeatletter

\@addtoreset{equation}{section}
\makeatother

\newcommand{\be}{\begin{eqnarray}}
  \newcommand{\ee}{\end{eqnarray}}
\newcommand{\ben}{\begin{eqnarray*}}
  \newcommand{\een}{\end{eqnarray*}}
\newcommand{\bec}{\begin{equation}\begin{array}{lll}}
    \newcommand{\eec}{\end{array}\end{equation}}
\newcommand{\nin}{\noindent}

\newcommand{\C}{\mathbb{C}}
\newcommand{\Z}{\mathbb{Z}}

\newcommand{\gl}{\lambda}

\newcommand{\ep}{\varepsilon}
\newcommand{\vphi}{\varphi}
\newcommand{\id}{\mathbb{1}}

\newcommand{\ot}{\otimes}
\newcommand{\sli}{\sum\limits}

\newcommand{\uq}{U_q(\widehat{\mathfrak{sl}}_2)}
\newcommand{\nn}{\nonumber}
\newcommand{\ds}{\displaystyle}

\newcommand{\mj}{\mathfrak{j}}
\newcommand{\bmj}{\bar{\mathfrak{j}}}
\newcommand{\bj}{\bar{j}}

\newcommand{\mref}[1]{(\ref{#1})}

\newcommand{\secref}[1]{Section~\ref{sec:#1}}

\newcommand{\vertex}[6]{
  \draw(0+#6,1) node[above] {#1};
  \draw(1+#6,0) node[right] {#2};
  \draw(0+#6,-1) node[below] {#3};
  \draw(-1+#6,0) node[left] {#4};
  \draw[aline=1] (0+#6,1) -- (0+#6,-1);
  \draw[aline=1] (1+#6,0) -- (-1+#6,0);
  \draw(0+#6,-2) node[below] {#5};}

\newcommand{\sqwt}[5]{\left(\left.\!\! \begin{array}{cc} #1 & #2 \\ #4 & #3 \end{array} \right| #5 \right)}
\newcommand{\trwt}[4]{\left(\left. #1 \, \genfrac{}{}{0pt}{}{#2}{#3}   \right| #4\right)}
\newcommand{\mat}[4]{\left( \begin{array}{cc} #1 & #2 \\ #3 & #4 \end{array}\right)}

\title{Conserved Currents in the Six-Vertex \\and Trigonometric Solid-On-Solid Models}

\author[1,2]{Yacine Ikhlef\thanks{\tt ikhlef@lpthe.jussieu.fr}}
\author[3]{Robert Weston\thanks{\tt R.A.Weston@hw.ac.uk}}
\affil[1]{Sorbonne Universit\'es, UPMC Univ Paris 06, UMR 7589, LPTHE, F-75005, Paris, France \bigskip}
\affil[2]{CNRS, UMR 7589, LPTHE, F-75005, Paris, France \bigskip}
\affil[3]{Department of Mathematics, Heriot-Watt University, Edinburgh EH14 4AS, UK,
  and Maxwell Institute for Mathematical Sciences, Edinburgh, UK}

\date{Dec. 12th,  2016}
\begin{document}
\maketitle

\bibliographystyle{unsrt}
\begin{abstract}
  \noindent 
 We construct quasi-local conserved currents in the six-vertex model with anisotropy parameter $\eta$ by making use of the quantum-group approach of Bernard and Felder. From these currents, we construct parafermionic operators with spin $1+i\eta/\pi$ that obey a discrete-integral condition around lattice plaquettes embedded into the complex plane. These operators are identified with primary fields in a $c=1$ compactified free Boson conformal field theory. We then consider a vertex-face correspondence that takes the six-vertex model to a trigonometric SOS model, and construct SOS operators that are the image of the six-vertex currents under this correspondence. We define corresponding SOS parafermionic operators with spins $s=1$ and $s=1+2i\eta/\pi$ that obey discrete integral conditions around SOS plaquettes embedded into the complex plane.  We consider in detail the cyclic-SOS case corresponding to the choice $\eta=i\pi (p-p')/p$, with $p'<p$ coprime. We identify our SOS parafermionic operators in terms of the screening operators and primary fields of the associated 
 $c=1-6(p-p')^2/pp'$ conformal field theory. 
\end{abstract}

\nopagebreak

\section{Introduction}
\label{sec:intro}

Conserved currents play a key role in the construction and analysis of quantum field theories. They are of particular interest in two dimensions: using complex co-ordinates $z=x+i t$ and $\bar z = x-it$, the current conservation law is
\begin{equation} \label{eq:h1}
  \partial_{\bar z} J - \partial_{z} \bar{J}=0 \,,
\end{equation}
and if the current obeys the additional condition
\begin{equation} \label{eq:h2}
 \partial_{\bar z} J + \partial_{z} \bar{J}=0 \,,
\end{equation}
then $\partial_{\bar z} J=0$ and so $J$ is a holomorphic field -- a hallmark of a conformal field theory (CFT). 

Suppose we consider instead a two-dimensional classical statistical mechanical model which at a given critical point is expected to correspond to a specific CFT. These correspondences are usually made by identifying critical indices, the central charge, finite-size behaviour, {\it etc}. Constructing and exploiting such connections has been one of the main activities and key successes of CFT since the earliest work \cite{BPZ84}. While useful and exhaustive, most of the work that deals with the critical continuum limit of 2D lattice models has been heuristic. Rigorous results exist only for a very restricted set of solvable models: dimers \cite{MR1872739} , the Ising model \cite{MR2680496,MR2957303}, critical percolation \cite{MR2227824,MR1851632}, and self-avoiding walks \cite{MR2912714}. A necessary requirement for the proof of theorems relating to the critical continuum limit has been the construction of an operator that obeys a discrete-lattice version of both conditions \eqref{eq:h1} and \eqref{eq:h2}. Such lattice operators are variously referred to as `discretely-holomorphic' or `preholomorphic' (see the review \cite{MR2827906} and references therein).

The search for discretely-holomorphic operators beyond the class of models related to the Ising model has led a number of authors to consider how such operators might be constructed for 2D solvable lattice models and their associated 1D quantum integrable systems \cite{RivaCardy,MR2441869,IkhlefRaj,IkhCardy,MR2942584,IWWZ,IkhWes16,MR3041918,MR2999732}. A systematic method of constructing currents obeying a discrete analogue of \eqref{eq:h1} for solvable models defined in terms of an underlying quantum group was formulated in \cite{BF91}. This approach leads naturally to quasi-local operators  -- meaning that the operators are associated with both a lattice insertion point and a path from this point to the lattice boundary. This quantum-group method has been previously exploited by the current authors and collaborators in order to construct quasi-local conserved currents for dense and dilute loop models\cite{IWWZ}  and for the chiral Potts model \cite{IkhWes16}. 

In general, the quasi-local conserved currents that come from the quantum-group approach do not, at least by construction, obey a discrete version of equation \eqref{eq:h2}; thus, they are not discretely holomorphic and their existence does not immediately open the way for a rigorous proof of the critical CFT limit.  Nevertheless they have other uses: their conformal spin naturally appears in a direct way (see for example \eqref{eq:pf1} below), and thus a heuristic identification with a CFT operator is almost immediate. Another, more formal, result is that the direct hexagonal-lattice analogue of the dense-loop-model operator was used in \cite{MR2912714} in the proof of a long-standing conjecture of Nienhuis regarding the connectivity constant of a hexagonal lattice \cite{MR751711,MR675241}. A further example  appeared in the context of the chiral Potts model in which the use of quantum group currents led to the definition of new parafermionic operators that generalise the parafermionic algebra of the $\mathbb{Z}_N$ clock model (which corresponds to the critical limit of the chiral Potts model) \cite{IkhWes16}. Furthermore, a careful analysis of the critical limit of the discrete version of \eqref{eq:h1} led to a perturbed CFT identification of the chiral Potts model that agreed with and extended a long-standing prediction of Cardy \cite{Cardy93}.

Motivated primarily by such practical uses, we describe in this paper the construction and CFT identification of conserved currents in the six-vertex (6V) model and a related trigonometric Solid-On-Solid (SOS) model. In the 6V model, these currents can be read off directly from the construction method of \cite{BF91}. By identifying the spin in the manner mentioned above, we connect the critical continuum limit of these operators with primary fields in a compactified free-boson CFT. In contrast, such a direct approach is not possible for the trigonometric SOS model, as in this case there is no underlying quantum group (there {\it is} a quasi-Hopf algebra \cite{MR1726695}, but this algebra lacks the required coalgebra structure that is central to the approach of \cite{BF91}). Instead, we rely on a vertex-face correspondence to map the currents of the 6V model to those of the SOS model. Given the quasi-local structure of the currents, it is by no means clear in advance that this mapping will lead to quasi-local operators in the SOS model, or that they will obey a discrete version of \eqref{eq:h1}. We show that both of these statements are true and go on to identify the operators in different CFT limits.

The content of this paper is as follows: in Section~\ref{sec:CR} we give a brief description of discrete versions of the Cauchy-Riemann relations. Then in Section~\ref{sec:6VSOS} we fix our conventions for the 6V and trigonometric SOS models, and give the vertex-face correspondence. In Section~\ref{sec:6Vcurrent}, we describe the construction of conserved currents within the algebraic picture of the 6V model, and identify the resulting parafermionic operators with primary fields of a compactified free boson CFT in the critical continuum limit.   In Section~\ref{sec:SOScurrent}, we map the 6V currents to SOS currents using the vertex-face correspondence, and define parafermionic operators in the SOS model. 
We then consider the special case of cyclic SOS models in Section~\ref{sec:CSOS}, and describe the identification of our lattice operators with local and parafermionic operators in CFT. Finally, we draw some conclusions in Section~\ref{sec:conclusion}.

\section{Discrete Current Conservation and Cauchy-Riemann Relations}
\label{sec:CR}

In this section, we explain the various forms of the discrete relations that we refer to in the rest of the paper. The Cauchy-Riemann relations for a complex function $J_x(x,t)+i J_t(x,t)$ are  
\begin{align}
  \partial_t J_t-\partial_x J_x &= 0 \,, \label{eq:rh1} \\
  \partial_x J_t +\partial_t J_x &= 0 \,. \label{eq:rh2}
\end{align}
Now consider a discretization of the above, where the functions are defined at the midpoints of a square unit-length lattice. More precisely, $J_x$ will be defined on horizontal edges only, and $J_t$ on vertical edges only. Then a possible discretization of these conditions is
\begin{align}
  J_t(\vec{u}_1) +J_x(\vec{u}_2)-J_t(\vec{u}_3)-J_x(\vec{u}_4) &= 0 \,, \label{eq:mdh1} \\
  J_x(\vec{v}_1) - J_t(\vec{v}_2)-J_x(\vec{v}_3)+J_t(\vec{v}_4) &= 0 \,, \label{eq:mdh2}
\end{align}
around the following plaquettes:
\begin{equation}
  \label{eq:plaquettes}
  \begin{tikzpicture}[scale=1,baseline=0]
    \draw (-1,0) -- (1,0);
    \draw(0,1) -- (0,-1);
    \draw (-1/2,0) node[below] {$\vec{u}_2$};
    \draw(-1/2,0) node {$*$};
    \draw (0,-1/2) node[right] {$\vec{u}_3$};
    \draw(0,-1/2) node {$*$};
    \draw (1/2,0) node[above] {$\vec{u}_4$};
    \draw(1/2,0) node {$*$};
    \draw (0,1/2) node[left] {$\vec{u}_1$};
    \draw(0,1/2) node {$*$};
  \end{tikzpicture}
  \qquad,\qquad
  \begin{tikzpicture}[scale=1,baseline=0]
    \draw (-0.5,-1) -- (-0.5,1);
    \draw (0.5,-1) -- (0.5,1);
    \draw (-1,0.5) -- (1,0.5);
    \draw (-1,-0.5) -- (1,-0.5);
    \draw (-1/2,0) node[left] {$\vec{v}_2$};
    \draw(-1/2,0) node {$*$};
    \draw (0,-1/2) node[below] {$\vec{v}_3$};
    \draw(0,-1/2) node {$*$};
    \draw (1/2,0) node[right] {$\vec{v}_4$};
    \draw(1/2,0) node {$*$};
    \draw (0,1/2) node[above] {$\vec{v}_1$};
    \draw(0,1/2) node {$*$};
  \end{tikzpicture}
  \quad.
\end{equation}
We can embed these relations into the complex plane in various ways.  The obvious way is to choose the embedding $\iota:(x,t)\mapsto x+it$. 
Then defining
$$
\Phi(x+it) =\begin{cases} 
J_x(x,t), & \quad \hbox{if $(x,t)$ is a horizontal edge,} \\
iJ_t(x,t), & \quad \hbox{if $(x,t)$ is a vertical edge,}
\end{cases}
$$
we see that the relations (\ref{eq:mdh1}--\ref{eq:mdh2}) become respectively
\begin{equation} \label{eq:plaqrelns}
  \sli_{i=1}^4 \Phi(z_i)\delta z_i=0, \qquad \sli_{i=1}^4 \Phi(w_i)\delta w_i=0 \,,
\end{equation}
where $z_i=\iota(\vec{u}_i)$,  $w_i=\iota(\vec{v}_i)$, and $\delta z_i$ and $\delta w_i$ are the anticlockwise oriented edges of the embedded plaquettes corresponding to \eqref{eq:plaquettes}. At this point, it is important to point out a potential source of confusion: we use the notation $\Phi(z)$  for this and other discrete functions in the complex plane (as opposed to, say, $\Phi(z,\bar z)$) but this does {\it not} imply that $\Phi$ is a holomorphic, or discretely holomorphic, function.

A more general embedding is $\iota:(x,t)\mapsto  e^{i \alpha_1}x+e^{i\alpha_2}t$, where $\alpha_1$ and $\alpha_2$ are the angles of the images of the $x$ and $t$ axes in the complex plane. Then we define
$$
  \Phi(e^{i \alpha_1} x+e^{i\alpha_2}t)=\begin{cases} 
    e^{i \alpha_1} J_x(x,t), & \quad \hbox{if $(x,t)$ is a horizontal edge,} \\
    e^{i \alpha_2} J_t(x,t), & \quad \hbox{if $(x,t)$ is a vertical edge,}
  \end{cases}
$$
The relations \eqref{eq:plaqrelns} still follow from the discrete relations (\ref{eq:mdh1}--\ref{eq:mdh2}) in this more general case. The sums in \eqref{eq:plaqrelns} may be interpreted as discrete integrals around the corresponding embedded rhombic plaquettes. 

 Thus, starting from just a current conservation law of the form \eqref{eq:mdh1}, we have different ways of choosing the embedding into the complex plane to obtain a discrete integral condition $\sum_{i=1}^4 \Phi(z_i)\delta z_i=0$ around a vertex. In the following sections, we shall use a particular embedding, whose choice is determined by consistency with the Yang-Baxter equation and crossing symmetry, and which produces an operator $\Phi(z)$ in a simple form with a prefactor that can be immediately interpreted in terms of the spin of the operator. 

\section{The Six-Vertex and Trigonometric SOS Models}
\label{sec:6VSOS}

In this section, we introduce our notation for the six-vertex model and describe a vertex-face correspondence between the Boltzmann weights of this model and those of a trigonometric SOS model. Our prescription is a trigonometric limit of the standard vertex-face correspondence between the eight-vertex and elliptic SOS models described in \cite{Baxter72ii,ABF84}. We were inspired to consider such trigonometric SOS models by the work \cite{MR2670981}. We will firstly recall how the partition function
of both models are related under the vertex-face correspondence. Then in Section~\ref{sec:SOScurrent}, we will go on to consider how correlation functions of our quasi-local operators behave under the vertex-face correspondence. 

\subsection{The six-vertex model}
We consider the six-vertex model with Boltzmann weights parametrized as 
\begin{center}
  \begin{tikzpicture}[scale=0.5]
    \vertex{$+$}{$+$}{$+$}{$+$}{$\sinh(\lambda+\eta)$}{0}
    \vertex{$-$}{$-$}{$-$}{$-$}{$\sinh(\lambda+\eta)$}{5}
    \vertex{$+$}{$-$}{$+$}{$-$}{$\sinh\lambda$}{10}
    \vertex{$-$}{$+$}{$-$}{$+$}{$\sinh\lambda$}{15}
    \vertex{$+$}{$-$}{$-$}{$+$}{$\sinh\eta$}{20}
    \vertex{$-$}{$+$}{$+$}{$-$}{$\sinh\eta$}{25}
  \end{tikzpicture}
\end{center}
where $\lambda$ and $\eta$ are complex numbers. Note that we use $\pm$ for the spin variables, while the arrows indicate the orientation of the lines of the lattice. The weights corresponding to the above vertices form the entries of the R-matrix 
$$
R(\lambda)^{\ep_1,\ep_2}_{\ep'_1,\ep'_2}=
\begin{tikzpicture}[scale=0.8,baseline=0]
  \vertex{$\ep_1$}{$\ep_2$}{$\ep'_1$}{$\ep'_2$}{}{0}
\end{tikzpicture} \quad,
$$
and it is sometimes useful to associate the spectral parameter $\lambda$ with the edges as follows:
$$
R(\lambda_1-\lambda_2)^{\ep_1,\ep_2}_{\ep'_1,\ep'_2}=
\begin{tikzpicture}[scale=0.8,baseline=0]
  \vertex{$\ep_1$}{$\ep_2$}{$\ep'_1$}{$\ep'_2$}{}{0};
  \draw(0.3,0.5) node[] {$\lambda_1$};
  \draw(0.5,-0.25) node[] {$\lambda_2$};
\end{tikzpicture} \quad.
$$
\nin In matrix-form we have
\begin{equation}\label{eq:Rmatrix}
  R(\lambda) = \left( \begin{array}{cccc}
      \sinh(\lambda+\eta) & 0 & 0 & 0 \\
      0 & \sinh\lambda & \sinh \eta & 0 \\
      0 & \sinh \eta & \sinh \lambda & 0 \\
      0 & 0 & 0 & \sinh(\lambda+\eta)
    \end{array} \right) \,.
\end{equation}

\noindent The R-matrix obeys the Yang-Baxter equation
\begin{equation} \label{eq:YBE-6V}
  R_{12}(\lambda_1-\lambda_2) R_{13}(\lambda_1-\lambda_3) R_{23}(\lambda_2-\lambda_3)
  = R_{23}(\lambda_2-\lambda_3) R_{13}(\lambda_1-\lambda_3) R_{12}(\lambda_1-\lambda_2) \,,
\end{equation}
which can be represented as:
$$
\begin{tikzpicture}[scale=0.6,baseline=0.3cm]
  \draw(-2,1) node[left] {$\lambda_1$};
  \draw(0,2.5) node[above] {$\lambda_2$};
  \draw(1,2) node[above] {$\lambda_3$};
  \draw[aline=1] (-2,1) -- (1,-1);
  \draw[aline=1] (0,2.5) -- (0,-1.5);
  \draw[aline=1] (1,2) -- (-2,0);
\end{tikzpicture}
\quad = \quad
\begin{tikzpicture}[scale=0.6,baseline=0.3cm]
  \draw(-1,2) node[above] {$\lambda_1$};
  \draw(0,2.5) node[above] {$\lambda_2$};
  \draw(2,1) node[right] {$\lambda_3$};
  \draw[aline=1] (2,1) -- (-1,-1);
  \draw[aline=1] (0,2.5) -- (0,-1.5);
  \draw[aline=1] (-1,2) -- (2,0);
\end{tikzpicture}
\quad.
$$
The R-matrix also satisfies the (normalised) unitarity and crossing symmetry relations
\begin{align}
  R_{21}(-\lambda)R_{12}(\lambda) &=\sinh(\lambda+\eta)\sinh(-\lambda+\eta)\ \id \,, \label{eq:unitarity} \\
  R(\lambda)^{\ep_1,\ep_2}_{\ep'_1,\ep'_2} &= (-1)^{\frac{\ep_1+\ep'_1}{2}} R(-\lambda-\eta)^{\ep_2,-\ep'_1}_{\ep'_2,-\ep_1} \,.
  \label{eq:crossing}
\end{align}

\subsection{The trigonometric SOS model}
\label{sec:SOS}
Baxter showed in \cite{Baxter72ii} that it was possible to relate the eight-vertex model to an elliptic SOS model -- the latter is a height model on a square lattice with Boltzmann weights associated with the faces. This correspondence  was generalised in subsequent work \cite{MR865759} and also re-expressed in the language of the dynamical Yang-Baxter equation \cite{MR1370676}. We do not need or discuss this dynamical approach in this paper. 
\subsubsection{The vertex-face correspondence}
\label{sec:VFC}
The starting point for our trigonometric vertex-face correspondence is to introduce certain vector-valued functions of the spectral parameter $\lambda$ that depend upon pairs $(a,b)$ of {\it height variables} living in $x_0+\mathbb{Z}$ (where $x_0 \in \mathbb{C}$ is an arbitrary reference point) with $|a-b|=1$. These functions are sometimes called {\it Baxter intertwiners}. More precisely, we have a column-vector valued function $\psi(a,b|\lambda)$ and row-vector valued function  $\psi^*(a,b|\lambda)$ given by
\begin{align}
  \def\arraystretch{1.8}
  \psi(a,a\pm 1|\lambda) &= \left[ \begin{array}{c} \exp\left( \frac{-\lambda \pm  a\eta}{2} \right) \\ \exp\left( \frac{+\lambda\mp a\eta}{2}\right) \end{array} \right] \,, \label{eq:psi} \\
  \qquad \psi^*(a,a\pm 1|\lambda) &= \frac{\pm 1}{2\sinh a\eta} \left[
    \exp\left( \frac{+\lambda\pm a\eta}{2} \right),
    \quad
    -\exp\left( \frac{-\lambda\mp a\eta}{2} \right)
  \right] \,. \label{eq:psid}
\end{align}
The following pictorial representation \cite{KKW} of the vector components $\psi(a,b|\lambda)_\ep$, $\psi^*(a,b|\lambda)_\ep$, $\ep\in\{\pm 1\}$, is very useful:
\begin{center}
  \begin{tikzpicture}[scale=0.3]
    \draw(-2,0) node[left] {$\psi(a,b|\lambda)_\ep = \quad$};
    \draw(2,0) node[right] {$\quad,$};
    \draw[gline] (-2,0) node[bblob]{} -- (2,0) node[bblob] {};
    \draw[aline=0.5] (0,0) -- (0,-2) node[below] {$\epsilon$};
    \draw(-2,0) node[above] {$a$};
    \draw(+2,0) node[above] {$b$};
    \draw(-1,0) node[below] {$\lambda$};
    \draw(18,0) node[left] {$\psi^*(a,b|\lambda)_\ep = \quad$};
    \draw(22,0) node[right] {$\quad.$};
    \draw[gline] (18,0) node[bblob]{} -- (22,0) node[bblob] {};
    \draw[aline=0.5] (20,2) node[above] {$\epsilon$} -- (20,0);
    \draw(18,0) node[above] {$a$};
    \draw(22,0) node[above] {$b$};
    \draw(20,1) node[left] {$\lambda$};
  \end{tikzpicture}
\end{center}
The six-vertex R-matrix and Baxter intertwiners obey the following relations
\begin{align}
  R(\lambda_{12}) \left[\psi(a,b|\lambda_1) \otimes \psi(b,c|\lambda_2) \right]
  &= \sum_d \left[\psi(d,c|\lambda_1) \otimes \psi(a,d|\lambda_2) \right] \ W\sqwt abcd{\lambda_{12}} \,, \label{eq:VIRF1} \\
  [\psi^*(d,c|\lambda_1) \otimes \psi^*(a,d|\lambda_2)] R(\lambda_{12})
  &= \sum_b W\sqwt abcd{\lambda_{12}} \left[ \psi^*(a,b|\lambda_1) \otimes \psi^*(b,c|\lambda_2) \right] \,, \label{eq:VIRF2}
\end{align}
where we introduce the notation $\lambda_{ij}=\lambda_i-\lambda_j$. Here the function $W$ depends on four height variables $(a,b,c,d)$ and is non-zero in the case when $|a-b|=|b-c|=|c-d|=|d-a|=1$. Explicitly, we have for any $a \in x_0+\mathbb{Z}$
\begin{equation} \label{eq:W}
  \begin{aligned}
    W\sqwt{a}{a\pm 1}{a\pm 2}{a\pm 1}{\lambda} &= \sinh(\lambda+\eta),\\
    W\sqwt{a}{a\pm 1}{a}{a\mp 1}{\lambda} &= \frac{\sinh\lambda\ \sinh[(a\pm 1)\eta]}{\sinh(a\eta)},\\
    W\sqwt{a}{a\pm 1}{a}{a\pm 1}{\lambda} &= \frac{\sinh\eta\ \sinh(a\eta\mp \lambda)}{\sinh(a\eta)} .
  \end{aligned}
\end{equation}
This function will form the Boltzmann weight of a configuration of four heights around a face, which we
represent by the picture  
\begin{center}
  \begin{tikzpicture}[scale=0.3]
    \draw(-2,0) node[left] {$W\sqwt abcd\lambda \quad=\qquad$};
    \draw(-2,+2) node[left] {$a$};
    \draw(+2,+2) node[right] {$b$};
    \draw(+2,-2) node[right] {$c$};
    \draw(-2,-2) node[left] {$d$};
    \draw(0,-1) node[above] {$\lambda$};
    \draw[gline] (-2,2) node[bblob] {} -- (2,2) node[bblob] {} -- (2,-2) node[bblob] {} -- (-2,-2) node[bblob] {} -- (-2,2);
    \draw[gline] (-1,2) -- (-2,1);
    \draw(3,0) node[right] {$.$};
  \end{tikzpicture}
\end{center}
We refer to the statistical-mechanical model so-defined as the trigonometric SOS model (or just SOS model). 
With these graphical conventions, the vertex-face correspondence relations (\ref{eq:VIRF1}--\ref{eq:VIRF2}) become 
\be
&&\raisebox{-1.7cm}{\begin{tikzpicture}[scale=0.3] 
    \draw[gline] (-2,2) node[bblob]{} -- (2,2) node[bblob]{} -- (2,-2) node[bblob]{};
    \draw[aline=1] (0,2) -- (0,-3) node[below]{$\lambda_1$};
    \draw[aline=1] (2,0) -- (-3,0) node[left]{$\lambda_2$};
    \draw(-2,2) node[above]{$a$};
    \draw(2,2) node[above]{$b$};
    \draw(2,-2) node[right]{$c$};
    \node at (6,-0.5){$\ds = \quad\sli_d$};
    \def\x{16};
    \draw(\x-2,+2) node[above] {$a$};
    \draw(\x+2,+2) node[above] {$b$};
    \draw(\x+2,-2) node[below] {$c$};
    \draw(\x-2,-2) node[below] {$d$};
    \draw(\x+0,-1) node[above] {$\lambda_{12}$};
    \draw[gline] (\x-2,2) node[bblob] {} -- (\x+2,2) node[bblob] {} -- (\x+2,-2) node[bblob] {} -- (\x-2,-2) node[bblob] {} -- (\x-2,2);
    \draw[gline] (\x-1,2) -- (\x-2,1);
    \draw[aline=1] (\x+0,-2) -- (\x+0,-4) node[below]{$\lambda_1$};
    \draw[aline=1] (\x-2,0) -- (\x-4,0) node[left]{$\lambda_2$};
    \node at (20,-0.5){$,$};
  \end{tikzpicture}}\label{pic:VIRF1}
\\ 
&&\raisebox{-1cm}{\begin{tikzpicture}[scale=0.3] 
    \draw(-8,0) node[] {and};
    \draw[gline] (-2,2) node[bblob]{} -- (-2,-2) node[bblob]{} -- (2,-2) node[bblob]{};
    \draw[aline=0.25] (0,2) node[above]{$\lambda_1$} -- (0,-2);
    \draw[aline=0.25] (2,0)  node[right]{$\lambda_2$} -- (-2,0);
    \draw(-2,2) node[above]{$a$};
    \draw(-2,-2) node[left]{$d$};
    \draw(2,-2) node[right]{$c$};
    \node at (8,-0.5){$\ds = \quad\sli_b$};
    \def\x{14};
    \draw(\x-2,+2) node[above] {$a$};
    \draw(\x+2,+2) node[above] {$b$};
    \draw(\x+2,-2) node[below] {$c$};
    \draw(\x-2,-2) node[below] {$d$};
    \draw(\x+0,-1) node[above] {$\lambda_{12}$};
    \draw[gline] (\x-2,2) node[bblob] {} -- (\x+2,2) node[bblob] {} -- (\x+2,-2) node[bblob] {} -- (\x-2,-2) node[bblob] {} -- (\x-2,2);
    \draw[gline] (\x-1,2) -- (\x-2,1);
    \draw[aline=0.5] (\x+0,4) node[above]{$\lambda_1$} -- (\x+0,2);
    \draw[aline=0.5] (\x+4,0) node[right]{$\lambda_2$} -- (\x+2,0);
    \node at (22,-0.5){$.$};
  \end{tikzpicture}}\label{pic:VIRF2}
\ee
\nin The Baxter intertwiners $\psi$ and $\psi^*$ also obey the following inversion relations:
\begin{subequations} \label{eq:inv}
  \begin{align}
    \psi^*(a,c|\lambda) \psi(a,b|\lambda) &= \delta_{bc} \,, \label{eq:inv-a} \\
    \sum_b \psi(a,b|\lambda) \psi^*(a,b|\lambda) &= \id \,, \label{eq:inv-b}  \\
    \psi'(c,a|\lambda)  \psi(b,a|\lambda)\  &= \delta_{bc} \label{eq:inv-c} \\ 
    \sum_b \psi(b,a|\lambda) \psi'(b,a|\lambda)  &=  \id \,. \label{eq:inv-d} 
  \end{align}
\end{subequations}
where, following \cite{LaPu,KKW}, we define 
\ben \psi'(a,b|\gl)= \frac{\sinh(a \eta)}{\sinh(b \eta)} \psi^*(a,b|\gl) e^{\eta \sigma^z}.\een
If we represent this modified intertwiner as
\ben 
\begin{tikzpicture}[scale=0.3]
  \draw(18,0) node[left] {$\psi'(a,b|\lambda) = \quad$};
  \draw(22,0) node[right] {$\quad,$};
  \draw[gline,dashed] (18,0) node[bblob]{} -- (22,0) node[bblob] {};
  \draw[aline=0.5] (20,2) node[above] {$\epsilon$} -- (20,0);
  \draw(18,0) node[above] {$a$};
  \draw(22,0) node[above] {$b$};
  \draw(20,1) node[left] {$\lambda$};
\end{tikzpicture}
\een
then the inversion relations \eqref{eq:inv} appear as  
\begin{center}
  \begin{tabular}{ccc}
    \begin{tikzpicture}[scale=0.3] 
      \draw[gline] (4,1) node[bblob]{} -- (0,1) node[bblob]{};
      \draw[gline] (4,-1) node[bblob]{} -- (0,-1) node[bblob]{}; -- (4,-2) node[bblob]{};
      \draw (4,1) node[right]{$b$};
      \draw (0,1) node[left]{$a$};
      \draw (0,-1) node[left]{$a$};
      \draw (4,-1) node[right]{$c$};
      \draw[aline=0.5] (2,1) -- (2,-1);
      \draw(2,0) node[right]{$\lambda$};
      
      \node at (7,0){$=\delta_{bc}$};

    \end{tikzpicture}
    &&\begin{tikzpicture}[scale=0.3] 
      \draw(-2,0) node[left]{$\ds \sum_b$};
      \draw(4,1) node[right] {$b$};
      \draw(4,-1) node[right] {$b$};
      \draw[gline] (4,1) node[bblob]{} -- (0,1) node[bblob]{};
      \draw[gline] (4,-1) node[bblob]{} -- (0,-1) node[bblob]{}; -- (4,-2) node[bblob]{};
      \draw (0,1) node[left]{$a$};
      \draw (0,-1) node[left]{$a$};
      \draw[aline=0.5] (2,3)  node[right] {$\lambda$} -- (2,1);
      \draw[aline=0.5] (2,-1) -- (2,-3) node[right]{$\lambda$};
      \node at (7,0){$=$};
      \draw[aline=0.5] (12,2) node[right]{$\lambda$} -- (12,-2);
    \end{tikzpicture} \\
    (a) && (b) \\ \\
    \begin{tikzpicture}[scale=0.3] 
      \draw[gline] (4,1) node[bblob]{} -- (0,1) node[bblob]{};
      \draw[gline,dashed] (4,-1) node[bblob]{} -- (0,-1) node[bblob]{}; -- (4,-2) node[bblob]{};
      \draw (0,2) node[left]{$b$};
      \draw (4,1) node[right]{$a$};
      \draw (4,-1) node[right]{$a$};
      \draw (0,-2) node[left]{$c$};
      \draw[aline=0.5] (2,1) -- (2,-1);
      \draw(2,0) node[left]{$\lambda$};
      \node at (7,0){$= \delta_{bc}$};
    \end{tikzpicture}
    &&  
    \begin{tikzpicture}[scale=0.3] 
      \draw(-2,0) node[left]{$\ds \sum_b $};
      \draw(4,1) node[right] {$a$};
      \draw(4,-1) node[right] {$a$};
      \draw[gline,dashed] (4,1) node[bblob]{} -- (0,1) node[bblob]{};
      \draw[gline] (4,-1) node[bblob]{} -- (0,-1) node[bblob]{}; -- (4,-2) node[bblob]{};
      \draw (0,1) node[left]{$b$};
      \draw (0,-1) node[left]{$b$};
      \draw[aline=0.5] (2,3)  node[right] {$\lambda$} -- (2,1);
      \draw[aline=0.5] (2,-1) -- (2,-3) node[right]{$\lambda$};
      \node at (7,0){$=$};
      \draw[aline=0.5] (12,2) node[right]{$\lambda$} -- (12,-2);
    \end{tikzpicture}\\
    (c) && (d)
  \end{tabular}
\end{center}
The SOS model weights obey the face version of the Yang-Baxter equation 
\begin{align}
  &\sum_g W\sqwt{f}{g}{d}{e}{\lambda_{12}} W\sqwt{a}{b}{g}{f}{\lambda_{13}} W\sqwt{b}{c}{d}{g}{\lambda_{23}} \nn \\
  &\qquad = \sum_g W\sqwt{a}{g}{e}{f}{\lambda_{23}} W\sqwt{g}{c}{d}{e}{\lambda_{13}} W\sqwt{a}{b}{c}{g}{\lambda_{12}} \, .
  \label{eq:YBE-SOS}
\end{align}
Graphically, this is represented as:
\begin{center}
  \begin{tabular}{cc}
    \begin{tikzpicture}[scale=0.3] 

      \draw[gline] (0,0) -- (-2,-4);
      \draw[gline] (0,0) -- (4,0);
      \draw[gline] node[bblob]{}(0,0) -- (-2,4);
      \draw[gline] (-4,0) node[bblob]{} -- (-2,4) node[bblob]{} -- (2,4) node[bblob]{} -- (4,0) node[bblob]{}
      -- (2,-4) node[bblob]{} -- (-2,-4) node[bblob]{} -- (-4,0);

      \draw[gline] (-3.4,1.25) -- (-3.4,-1.25);
      \draw[gline] (-1,4) -- (-1.5,3);
      \draw[gline] (-1,-4) -- (-1.5,-3);

      \draw(-4,0) node[left] {$\ds \sum\limits_g^{\phantom{a^2}} \quad a$};
      \draw(-2,4) node[above] {$b$};
      \draw(2,4) node[above] {$c$};
      \draw(4,0) node[right] {$d$};
      \draw(2,-4) node[below] {$e$};
      \draw(-2,-4) node[below] {$f$};
      \draw(0,0) node[left] {$g$};
      \node at (-2,0) {$\lambda_{13}$};
      \node at (1,2) {$\lambda_{23}$};
      \node at (1,-2) {$\lambda_{12}$};

    \end{tikzpicture}

    \begin{tikzpicture}[scale=0.3] 

      \draw[gline] (0,0) -- (2,-4);
      \draw[gline] (0,0) -- (-4,0);
      \draw[gline] node[bblob]{}(0,0) -- (2,4);
      \draw[gline] (2,4) node[bblob]{} -- (-2,4) node[bblob]{} -- (-4,0) node[bblob]{}
      -- (-2,-4) node[bblob]{} -- (2,-4) node[bblob]{} -- (4,0) node[bblob]{} -- (2,4);

      \draw[gline] (0.6,1.25) -- (0.6,-1.25);
      \draw[gline] (-3,0) -- (-3.5,1);
      \draw[gline] (-3,0) -- (-3.5,-1);

      \draw(-4,0) node[left] {$\ds \quad = \qquad \sum\limits_g^{\phantom{a^2}} \quad a$};
      \draw(-2,4) node[above] {$b$};
      \draw(2,4) node[above] {$c$};
      \draw(4,0) node[right] {$d \qquad.$};
      \draw(2,-4) node[below] {$e$};
      \draw(-2,-4) node[below] {$f$};
      \draw(0,0) node[above] {$g\ $};
      \node at (2,0) {$\lambda_{13}$};
      \node at (-1,2) {$\lambda_{12}$};
      \node at (-1,-2) {$\lambda_{23}$};

    \end{tikzpicture}
  \end{tabular}
\end{center}

\subsubsection{Correspondence of the partition functions}

The correspondence between the partition functions of the six-vertex and SOS models starts by dressing the
boundary of the six-vertex model with the Baxter intertwiners, with the external boundary heights fixed. 
The vertex-face correspondence is used repeatedly starting either from the NE or SW corner of the
lattice. If we start from the NE corner and use \eqref{eq:VIRF1} then we arrive at the correspondence
$$
\begin{tikzpicture}[scale=0.7,baseline=2.1cm]
  \foreach\x in {1,2,3,4,5,6} 
  {
    \draw[aline=0.2] (\x,6.5) -- (\x,0.5);
    \draw(\x-0.5,0.5) node[bblob] {};
    \draw(\x-0.5,6.5) node[bblob] {};
  }
  \foreach\y in {1,2,3,4,5,6}{
    \draw[aline=0.2] (6.5,\y) -- (0.5,\y);
    \draw(0.5,\y+0.5) node[bblob] {};
    \draw(6.5,\y+0.5) node[bblob] {};
  }
  \draw(6.5,0.5) node[bblob] {};
  \draw[gline] (0.5,0.5) -- (6.5,0.5) -- (6.5,6.5) -- (0.5,6.5) -- (0.5,0.5);
\end{tikzpicture}
\quad = \quad
\begin{tikzpicture}[scale=0.7,baseline=2.1cm]
  \foreach\x in {1,2,3,4,5,6}{
    \foreach\y in {1,2,3,4,5,6}{
      \sosplaq(\x,\y);
    }
  }
  \foreach\x in {2,3,4,5,6}{
    \draw(\x-0.5,-0.2) node[bblob]{};
    \draw(-0.2,\x-0.5) node[bblob]{};
    \draw[aline=0.5] (0.5,\x-1) -- (-0.2,\x-1);
    \draw[aline=0.5] (\x-1,0.5) -- (\x-1,-0.2);
  }
  \draw(-0.2,-0.2)  node[bblob]{};
  \draw[aline=0.8] (0.5,6) -- (0.18,6);
  \draw[aline=0.8] (6,0.5) -- (6,0.18);
  \draw[gline] (6.5,0.5)-- (5.5,-0.2) -- (-0.2,-0.2) -- (-0.2,5.5) -- (0.5,6.5);
\end{tikzpicture}
\quad ,
$$
where all internal height variables are summed over. We can then apply the relation \eqref{eq:inv-a} repeatedly starting from the NW corner, in order to arrive at equality with the pure SOS partition function: \vspace*{4mm}
$$
\begin{tikzpicture}[scale=0.7,baseline=2.1cm]
  \foreach\x in {1,2,3,4,5,6}{
    \foreach\y in {1,2,3,4,5,6}{
      \sosplaq(\x,\y);
    }
  }
\end{tikzpicture}
\quad.
$$
\bigskip

We can also apply this vertex-face correspondence to connect correlation functions of local operators in the six-vertex model to those in the SOS model. This approach was used successfully to compute such correlation functions in the eight-vertex case in the thermodynamic limit in \cite{LaPu,KKW}. In this paper we are interested in using the vertex-face correspondence to connect correlation functions of {\it quasi-local} operators in the six-vertex and SOS models for finite-size lattices. We shall introduce the relevant six-vertex model quasi-local operators in the next section.

\section{Conserved Currents in the Six-Vertex Model}
\label{sec:6Vcurrent}

In this section, we recall the relation of the six-vertex model to the quantum affine algebra $\uq$, and construct quasi-local conserved currents in terms of the Chevalley generators of this algebra. This way of relating quantum groups and quasi-local conserved currents was proposed in a more general setting by Bernard and Felder \cite{BF91} and was exploited in the context of discrete holomorphicity in previous work of the authors and collaborators \cite{IWWZ,IkhWes16}.

\subsection{The algebraic picture of the six-vertex model}
The algebra $\uq$ is an associative algebra generated by the Chevalley generators $e_i,f_i,t_i^{\pm 1}$, $i\in\{0,1\}$, with 
relations
\begin{equation}
  t_i e_j t_i^{-1}=q^{A_{ij}} e_j \,,
  \quad  t_i f_j t_i^{-1}=q^{-A_{ij}} f_j \,,
  \quad [e_i,f_j]=\delta_{i,j} \frac{t_i-t_i^{-1}}{q-q^{-1}} \,,
  \quad A=\begin{pmatrix} 2&-2\\-2 &2\end{pmatrix} \,,
\end{equation}
and additional Serre relations (a gentle introduction to quantum affine algebras can be found in \cite{MR1239667}). Here $q$ is a parameter, which we can take to be a complex number, and we write:
$$
q = e^\eta \,.
$$
The algebra is a Hopf algebra, meaning that it has additional structure including a coproduct $\Delta:\uq  \rightarrow \uq\otimes \uq$ and an antipode $S:\uq  \rightarrow \uq$, which we here choose as 
\begin{align}
  & \Delta(e_i)=e_i\ot \id + t_i\ot e_i \,,
  \quad && \Delta(f_i)=f_i\ot t_i^{-1} + \id \ot f_i \,,
  \quad && \Delta(t_i^{\pm 1})=t_i^{\pm 1} \ot t_i^{\pm 1}, \label{eq:Delta1} \\
  & S(e_i)= -t_i^{-1} e_i \,,
  \quad && S(f_i)= -f_i t_i \,,
  \quad && S(t_i^{\pm 1})=t_i^{\mp 1} \,.
\end{align}

The representation of $\uq$ that is relevant to the six-vertex model is that specified by the 
module $V_\lambda=\C^2\ot \C[e^{\lambda},e^{-\lambda}]$ with action
\begin{equation} \label{eq:Vlambda}
  \begin{aligned}
    e_1 \mapsto \mat{0}{e^{\lambda}}{0}{0} \,,
    \qquad  f_1 \mapsto \mat{0}{0}{e^{-\lambda}}{0} \,,
    \qquad t_1 \mapsto \mat{e^{\eta}}{0}{0}{e^{-\eta}} \,, \\
    e_0 \mapsto \mat{0}{0}{e^{\lambda}}{0} \,,
    \qquad f_0 \mapsto \mat{0}{e^{-\lambda}}{0}{0} \,,
    \qquad t_0 \mapsto \mat{e^{-\eta}}{0}{0}{e^{\eta}} \,.
  \end{aligned} 
\end{equation}
 The R-matrix \eqref{eq:Rmatrix} is then the map
$$R(\lambda_{1}-\lambda_2): V_{\lambda_1}\ot  V_{\lambda_2}\rightarrow V_{\lambda_1}\ot  V_{\lambda_2}$$
such that, for any $x \in \uq$,
\begin{equation} \label{eq:rdef}
  R(\lambda_{1}-\lambda_2) \circ \Delta(x)= \Delta'(x)  \circ R(\lambda_{1}-\lambda_2) \,,
\end{equation}
where $\Delta'(x)=P \circ \Delta(x) \circ P $ is the opposite co-product with $P(x\otimes y) = y\otimes x$.

\subsection{Currents corresponding to Chevalley generators}
\label{sec:Chev}

For our purposes, it is useful to consider the generators $\bar f_i=  e_i\, t_i^{-1}$ with 
\begin{equation} \label{eq:Delta2}
  \Delta(\bar f_i)= \bar f_i \ot t_i^{-1} + \id \ot \bar f_j \,,
\end{equation}
and
\begin{equation} \label{eq:fbarrep}
  \bar f_1 \mapsto \mat{0}{e^{\lambda+\eta}}{0}{0}, \quad \bar f_0 \mapsto \mat{0}{0}{e^{\lambda+\eta}}{0} \,,
\end{equation}
following from the above. The coproduct structure means that there is action of $\uq$ on the $N$-fold tensor product $V_\gl^{\ot N}$ given by the iterated coproduct $\Delta^{(N)}: \uq \to \uq^{\ot N}$, with $\Delta^{(N)}=(\Delta \ot \id ) \Delta^{(N-1)}$ and $\Delta^{(2)}=\Delta$. The action of $f_i$, $\bar f_i$ and $t_i^{-1}$ on this space is given by
\ben 
&&\Delta^{(N)}(f_i)=\sli_{j=1}^N \id \ot \id \ot \cdots \id\ot \!\!\!\!\stackrel[(j\mathrm{-th})]{}{f_i}\!\! \!\!\ot t_i^{-1}\ot t_i^{-1} \ot \cdots t_i^{-1},\\
&&\Delta^{(N)}(\bar f_i)=\sli_{j=1}^N \id \ot \id \ot \cdots \id\ot \!\!\!\!\stackrel[(j\mathrm{-th})]{}{\bar f_i} \!\!\!\!\ot t_i^{-1}\ot t_i^{-1} \ot \cdots t_i^{-1},\\
&&\Delta^{(N)}(t_i^{\pm 1})= t_i^{\pm 1} \ot  t_i^{\pm 1} \ot \cdots \ot t_i^{\pm 1} \ot  t_i^{\pm 1}.
\een

Now we adopt the pictorial representation of \cite{BF91}: we represent the space $V_{\gl}$ by a downwards directed edge
and the action of either $f_i$ or $\bar f_i$ as\footnote{We could distinguish the operators $f_i$ or $\bar f_i$ pictorially by attaching a $i$ or $\bar i$ index to the wavy line,  but here we suppress this additional information in order to simplify the notation.}
\begin{equation*}
  f_i , \bar f_i =
  \quad \begin{tikzpicture}[baseline=0]
    \draw[aline=0.9] (0,0.5) -- (0,-0.5);
    \draw[blue,wavy] (0.7,0) -- (0,0);
    \cross(0,0);
  \end{tikzpicture}
  \quad.
\end{equation*}
Furthermore, we represent the action of $t_i$ and $t_i^{-1}$ on $V_\gl$ by
\begin{equation*}
  t_i=
  \quad\raisebox{-2.5ex}{\begin{tikzpicture}
      \draw[aline=0.9] (0,0.5) -- (0,-0.5);
      \draw[blue,wavy=1] (-0.5,0) -- (0.5,0);
    \end{tikzpicture}}
  \quad,\qquad 
  t_i^{-1}=
  \quad\raisebox{-2.5ex}{\begin{tikzpicture}
      \draw[aline=0.9] (0,0.5) -- (0,-0.5);
      \draw[blue,wavy=1] (0.5,0) -- (-0.5,0);
    \end{tikzpicture}} \quad.
\end{equation*}
The action of $f_i$ or $\bar f_i$ on the space $V_\gl^{\ot N}$ can be represented by
\begin{equation*}
  \sum_{j=1}^N
  \quad \raisebox{-5.5ex}{\begin{tikzpicture}
      \foreach\x in {1,2,3,4,5,6}{
        \draw[aline=0.9] (\x,0.5) -- (\x,-0.5);
      }
      \cross(3,0);
      \draw[blue,wavy=0.5] (6.5,0) -- (3,0);
      \draw(3,-0.8) node[] {($j$-th)};
    \end{tikzpicture}} \quad,
\end{equation*}
and the action of $t_i$ and $ t^{-1}_i$ by
\begin{align*}
  t_i &=
  \quad \raisebox{-2.5ex}{\begin{tikzpicture}
      \foreach\x in {1,2,3,4,5,6}{
        \draw[aline=0.9] (\x,0.5) -- (\x,-0.5);
      };
      \draw[blue,wavy=0.5] (0.5,0) -- (6.5,0);
    \end{tikzpicture}} \quad, \\ \\
  t_i^{-1} &=
  \quad \raisebox{-2.5ex}{\begin{tikzpicture}
      \foreach\x in {1,2,3,4,5,6}{
        \draw[aline=0.9] (\x,0.5) -- (\x,-0.5);
      };
      \draw[blue,wavy=0.5] (6.5,0) -- (0.5,0);
    \end{tikzpicture}} \quad.
\end{align*}

\subsection{Elementary moves of tail operators}

The (negative of) antipodes $-S(f_i)=f_i t_i=:\bar e_i$ and $-S(\bar f_i)= \bar f_i t_i=e_i$ then have the natural graphical realisation
\begin{center}
  \begin{tikzpicture}[baseline=0]
    \draw[aline=0.9] (0,1) -- (0,-0.5);
    \draw[blue,wavy] (-0.7,0.5) -- (0.5,0.5) -- (0.5,0) -- (0,0);
    \cross(0,0);
  \end{tikzpicture} \quad,
\end{center}
which we can in turn abbreviate by 
\begin{center}
  \begin{tikzpicture}[baseline=0]
    \draw[aline=0.9] (0,0.5) -- (0,-0.5);
    \draw[blue,wavy] (-0.7,0) -- (0,0);
    \cross(0,0);
  \end{tikzpicture} \quad.
\end{center}
The relations $t_i f_i t_i^{-1}=e^{-2\eta } f_i$ and $t_i^{-1} f_i t_i=e^{2\eta } f_i$ (and similar relations with $f_i\rightarrow \bar f_i$), appear as the following:
\begin{equation} \label{eq:winding}
  \begin{tikzpicture}[baseline=0]
    \draw[aline=0.9] (0,1) -- (0,-1.0);
    \draw[blue,wavy] (0.7,0.5) -- (-0.5,0.5) -- (-0.5,-0.5) -- (0.5,-0.5) -- (0.5,0) -- (0,0);
    \cross(0,0);
  \end{tikzpicture}
  \quad = e^{-2\eta} \quad
  \begin{tikzpicture}[baseline=0]
    \draw[aline=0.9] (0,1) -- (0,-1.0);
    \draw[blue,wavy] (0.7,0) --  (0,0);
    \cross(0,0);
  \end{tikzpicture}
  \quad,\qquad\qquad
  \begin{tikzpicture}[baseline=0]
    \draw[aline=0.9] (0,1) -- (0,-1.0);
    \draw[blue,wavy] (0.7,-0.5) -- (-0.5,-0.5) -- (-0.5,0.5) -- (0.5,0.5) -- (0.5,0) -- (0,0);
    \cross(0,0);
  \end{tikzpicture}
  \quad = e^{2\eta} \quad
  \begin{tikzpicture}[baseline=0]
    \draw[aline=0.9] (0,1) -- (0,-1.0);
    \draw[blue,wavy] (0.7,0) --  (0,0);
    \cross(0,0);
  \end{tikzpicture}
  \quad.
\end{equation}
Finally, we note that $t_i^{-1} t_i=\id = t_i t_i^{-1}$ has the diagrammatic realisation
\begin{equation}
  \begin{tikzpicture}[scale=0.7,baseline=0]
    \draw[aline=1] (0,1) -- (0,-1);
    \draw[blue,wavy=0.5] (-0.5,1) -- (-0.5,-1);
  \end{tikzpicture}
  \quad=\quad
  \begin{tikzpicture}[scale=0.7,baseline=0]
    \draw[aline=1] (0,1) -- (0,-1);
    \draw[blue,wavy=0.5] (-0.5,1) -- (-0.5,0.5) -- (0.5,0.5) -- (0.5,-0.5) -- (-0.5,-0.5) -- (-0.5,-1);
  \end{tikzpicture}
  \qquad,\qquad
  \begin{tikzpicture}[scale=0.7,baseline=0]
    \draw[aline=1] (0,1) -- (0,-1);
    \draw[blue,wavy=0.5] (0.5,1) -- (0.5,-1);
  \end{tikzpicture}
  \quad=\quad
  \begin{tikzpicture}[scale=0.7,baseline=0]
    \draw[aline=1] (0,1) -- (0,-1);
    \draw[blue,wavy=0.5] (0.5,1) -- (0.5,0.5) -- (-0.5,0.5) -- (-0.5,-0.5) -- (0.5,-0.5) -- (0.5,-1);
  \end{tikzpicture}
  \quad.
\end{equation}
Using \eqref{eq:rdef} with $x=t_i^{\pm 1}$ we have
\begin{align}
  \begin{tikzpicture}[baseline=0]
    \draw[aline=0.98] (0,0.7) -- (0,-0.7);
    \draw[aline=0.98] (0.7,0) -- (-0.7,0);
    \draw[blue,wavy=0.5] (0.7,-0.35) --(-0.35,0.7);
  \end{tikzpicture}
  \quad &= \quad
  \begin{tikzpicture}[baseline=0]
    \draw[aline=0.98] (0,0.7) -- (0,-0.7);
    \draw[aline=0.98] (0.7,0) -- (-0.7,0);
    \draw[blue,wavy=0.5] (0.35,-0.7) --(-0.7,0.35);
  \end{tikzpicture}
  \quad, \label{eq:tailmove1} \\
  \begin{tikzpicture}[baseline=0]
    \draw[aline=0.98] (0,0.7) -- (0,-0.7);
    \draw[aline=0.98] (0.7,0) -- (-0.7,0);
    \draw[blue,wavy=0.5](-0.35,0.7)-- (0.7,-0.35);
  \end{tikzpicture}\quad
  \quad &= \quad
  \begin{tikzpicture}[baseline=0]
    \draw[aline=0.98] (0,0.7) -- (0,-0.7);
    \draw[aline=0.98] (0.7,0) -- (-0.7,0);
    \draw[blue,wavy=0.5] (-0.7,0.35) -- (0.35,-0.7);
  \end{tikzpicture}
  \quad. \label{eq:tailmove2}
\end{align}

\subsection{Current conservation}

Now let us use the R-matrix commutation relation \eqref{eq:rdef}  with $x=f_i,\bar f_i$. Noting the coproduct formulae \eqref{eq:Delta1} and \eqref{eq:Delta2},  it is apparent that this commutation relation becomes  
\begin{equation} \label{eq:fint}
  \begin{tikzpicture}[baseline=0]
    \draw[aline=0.98] (0,0.7) -- (0,-0.7);
    \draw[aline=0.98] (0.7,0) -- (-0.7,0);
    \cross(0,0.35);
    \draw[blue,wavy=0.5] (0.7,-0.35) --(0.35,-0.35) -- (0.35,0.35) -- (0,0.35);
  \end{tikzpicture}
  \quad + \quad
  \begin{tikzpicture}[baseline=0]
    \draw[aline=0.98] (0,0.7) -- (0,-0.7);
    \draw[aline=0.98] (0.7,0) -- (-0.7,0);
    \cross(0.35,0);
    \draw[blue,wavy=0.5] (0.7,-0.35) --(0.35,-0.35) -- (0.35,0);
  \end{tikzpicture}
  \quad = \quad
  \begin{tikzpicture}[baseline=0]
    \draw[aline=0.98] (0,0.7) -- (0,-0.7);
    \draw[aline=0.98] (0.7,0) -- (-0.7,0);
    \cross(-0.35,0);
    \draw[blue,wavy=0.3] (0.7,-0.35) --(-0.35,-0.35) -- (-0.35,0);
  \end{tikzpicture}
  \quad + \quad
  \begin{tikzpicture}[baseline=0]
    \draw[aline=0.98] (0,0.7) -- (0,-0.7);
    \draw[aline=0.98] (0.7,0) -- (-0.7,0);
    \cross(0,-0.35);
    \draw[blue,wavy=0.3] (0.7,-0.35) --(0,-0.35);
  \end{tikzpicture}
  \quad.
\end{equation}
Hence, we have the four-term relation 
\ben
&&\quad\quad\begin{tikzpicture}
  \foreach\x in {1,2,3,4,5,6}{
    \draw[aline=0.95] (\x,0.7) -- (\x,-0.7);
  }
  \draw[aline=0.98] (6.5,0) -- (0.5,0);
  \cross(3,0.35);
  \draw[blue,wavy=0.5] (6.5,-0.25) --(3.5,-0.25) -- (3.25,0.35) -- (3,0.35);
  \draw(7,0) node[] {$+$};
\end{tikzpicture}\quad 
\begin{tikzpicture}
  \foreach\x in {1,2,3,4,5,6}{
    \draw[aline=0.9] (\x,0.7) -- (\x,-0.7);
  }
  \draw[aline=0.98] (6.5,0) -- (0.5,0);
  \cross(3.5,0);
  \draw[blue,wavy=0.6] (6.5,-0.25) --(3.5,-0.25) -- (3.5,0);
\end{tikzpicture}\\[3mm]
&& \begin{tikzpicture}\draw(0,0) node[] {$-$};
  \foreach\x in {1,2,3,4,5,6}{
    \draw[aline=0.9] (\x,0.7) -- (\x,-0.7);
  }
  \draw[aline=0.98] (6.5,0) -- (0.5,0);
  \cross(2.5,0);
  \draw[blue,wavy=0.5] (6.5,-0.25) -- (2.5,-0.25) -- (2.5,0);
  \draw(7,0) node[] {$-$};
\end{tikzpicture}\quad 
\begin{tikzpicture}
  \foreach\x in {1,2,3,4,5,6}{
    \draw[aline=0.9] (\x,0.7) -- (\x,-0.7);
  }
  \draw[aline=0.98] (6.5,0) -- (0.5,0);
  \cross(3,-0.25);
  \draw[blue,wavy=0.5] (6.5,-0.25) -- (3,-0.25);
  \draw(7,0) node[] {$=0$};
\end{tikzpicture}\een
around any vertex (the third from the left in the example) in a product of R-matrices.
Also note that using \eqref{eq:tailmove1} and \eqref{eq:tailmove2} one
can move the wavy `tail' corresponding to either $t_i$ or $t_i^{-1}$ through any vertex. Exploiting this
ability we find for example that
$$
\begin{tikzpicture}[baseline=0]
  \foreach\x in {1,2,3,4,5,6}{
    \draw[aline=0.95] (\x,0.7) -- (\x,-0.7);
  }
  \draw[aline=0.98] (6.5,0) -- (0.5,0);
  \cross(3,0.35);
  \draw[blue,wavy=0.5] (6.5,-0.25) --(3.5,-0.25) -- (3.25,0.35) -- (3,0.35);
\end{tikzpicture}
\quad = \quad
\begin{tikzpicture}[baseline=0]
  \foreach\x in {1,2,3,4,5,6}{
    \draw[aline=0.9] (\x,0.7) -- (\x,-0.7);
  }
  \draw[aline=0.98] (6.5,0) -- (0.5,0);
  \cross(3,0.35);
  \draw[blue,wavy=0.6] (6.5,-0.25) -- (6.25,-0.25) -- (6.25,0.35) -- (3,0.35);
\end{tikzpicture} \quad.
$$

We can extend these various relations to the computation of correlation functions of these quasi-local operator insertions into the 2D lattice. For example we might compute the correlation function corresponding to the configuration sum with the following operator insertion at the lattice point $\vec{r}$ (corresponding to the cross):
\begin{center}
  \begin{tikzpicture}[scale=0.7,baseline=2.1cm]
    \foreach\x in {1,2,3,4,5,6} 
    {
      \draw[aline=0.99] (\x,6.5) -- (\x,0.5);
    }
    \foreach\y in {1,2,3,4,5,6}{
      \draw[aline=0.99] (6.5,\y) -- (0.5,\y);
    }
    \draw[blue,wavy=0.4] (5.5,0.5) -- (5.5,3.5);
    \draw[blue,wavy=0.4] (5.5,3.5) -- (3.5,3.5) -- (3.5,4.5) -- (2,4.5);
    \cross(2,4.5);
  \end{tikzpicture}
  \quad.
\end{center}
Dividing this configuration sum by the partition function, we obtain a quasi-local current expectation value, which we can denote by $\langle j(\vec{r})\rangle$. It will depend on the insertion point $\vec{r}$ and the boundary point (which we assume fixed), but not on the precise path followed by the tail -- which can be moved according to (\ref{eq:tailmove1}--\ref{eq:tailmove2}). If we denote the four lattice insertion points around a vertex as $(\vec{r}_1,\vec{r}_2,\vec{r}_3,\vec{r}_4)$ corresponding to the picture
\begin{equation}
  \label{fig:vplaq}
  \begin{tikzpicture}[baseline=0]
    \draw[aline=0.98] (1,0) -- (-1,0);
    \draw[aline=0.98] (0,1) -- (0,-1);
    \draw(0,0.5) node[] {$*$};
    \draw(0,-0.5) node[] {$*$};
    \draw(0.5,0) node[] {$*$};
    \draw(-0.5,0) node[] {$*$};
    \draw(0,0.5) node[left] {$\vec{r}_1$};
    \draw(0,-0.5) node[right] {$\vec{r}_3$};
    \draw(0.5,0) node[above] {$\vec{r}_4$};
    \draw(-0.5,0) node[below] {$\vec{r}_2$};
  \end{tikzpicture}
  \quad,
\end{equation}
then the four-term relation discussed above will be of the form 
\begin{equation} \label{eq:currentreln}
  j(\vec{r}_1)-  j(\vec{r}_2)-  j(\vec{r}_3)+  j(\vec{r}_4)=0 \,,
\end{equation}
where we have dropped the implied expectation values to simplify notation. 

Suppose now that we have a quasi-local operator of the above form for which the tail has a non-zero winding 
number $M$ around the insertion point. Using the relations \eqref{eq:winding}, \eqref{eq:tailmove1} and \eqref{eq:tailmove2}, we can `unwind' the tail at the expense of picking up a factor $e^{-2\eta M}$.

\subsection{Parafermionic operators}
\label{sec:para-6v}

We now follow the procedure, outlined in Section~\ref{sec:CR}, in which we embed the 2D lattice into the complex plane and interpret the relation \eqref{eq:currentreln} as a discrete integral condition around a rhombus\footnote{Note that there are slight modifications to the discussion in Section 2 resulting from the choice of orientation of our R-matrices.}. We will embed the lattice into the plane so that the plaquette \eqref{fig:vplaq} becomes a unit rhombus with angle $\theta$ thus:
\be
\begin{tikzpicture}[scale=0.5,baseline=2]
  \draw[gline] (-4,0) node[bblob]{} -- (0,3) node[bblob]{} -- (4,0)
node[bblob]{} -- (0,-3) node[bblob]{} -- (-4,0);
\draw (-2,1.5) .. controls (-0.8,0) and (0.8,0) .. (2,-1.5);
\draw[aline] (2,-1.5) --(3,-4+3.5/3);
\draw (-2,1.5) --(-3,4-3.5/3) node[above] {$\lambda_1$};
\draw (2,1.5) .. controls (0.8,0) and (-0.8,0) .. (-2,-1.5);
\draw[aline] (-2,-1.5) --(-3,-4+3.5/3);
\draw (2,1.5) --(3,4-3.5/3) node[above] {$\lambda_2$};;
\draw (-3,9/4-3) arc (-36.9:36.9:1.2);
\draw (-3.3,0) node[] {$\theta$};
  \draw(-2,1.5) node[] {$*$};
  \draw(-2,-1.5) node[] {$*$};
  \draw(2,1.5) node[] {$*$};
  \draw(2,-1.5) node[] {$*$};
   \draw(-2.2,1.5) node[left] {$z_1$};
   \draw(-2.2,-1.5) node[left] {$z_2$};
   \draw(2.2,-1.5) node[right] {$z_3$};
   \draw(2.2,1.5) node[right] {$z_4$};
\end{tikzpicture}
\label{fig:embed}
\ee 
where $z_i\in \C$ denotes the image of $\vec{r}_i$.
This embedding is achieved by mapping a line with spectral parameter $\lambda$ to a line in the complex plane characterised by the angle $\alpha$ defined by 
\begin{equation}
  \label{fig:embed2}
  \begin{tikzpicture}[scale=2,baseline=0]
    \draw[gline] (-1,0)  -- (0,1);
    \draw[aline=1] (-1,1) node[above]{$\lambda$} -- (0,0);
    \draw[dashed] (-0.5,0.5) -- (0.5,0.5);
    \draw (0,0.5) arc (0:45:0.5);
    \draw (-0.25,0.6) node {$\alpha$};
  \end{tikzpicture}
  \quad,
\end{equation}
where the green and black lines meet at 90 degrees, and in which $\alpha$ is specified in terms of $\lambda$ by the relation
\begin{equation} \label{eq:alphalambda}
  \alpha=-\frac{\pi \lambda}{\eta} \,.
\end{equation}
With these conventions, we have that $\theta$ in \eqref{fig:embed} is given by
\be \label{eq:embed}
\theta = \alpha_1-\alpha_2=\frac{\pi (\lambda_2-\lambda_1)}{\eta}.
\ee
which is consistent with both the Yang-Baxter relation \eqref{eq:YBE-6V} (interpreted as a geometrical relation in the plane) and the crossing relation \eqref{eq:crossing} (interpreted as reversing the direction of a line). Note that in order that in order that the intersection point of black and green lines occurs at the midpoints of the edges of the rhombus, we need to distort the black lines as shown in \eqref{fig:embed}.

We now need to introduce two more pieces of notation: firstly, we use $\alpha(z)$ to refer to the angle of the line passing through an embedded point $z$ thus:
\begin{equation}
  \begin{tikzpicture}[scale=2,baseline=0]
    \draw[gline] (-1,0)  -- (0,1);
    \draw[aline=1] (-1,1)  -- (0,0);
    \draw[dashed] (-0.5,0.5) -- (0.5,0.5);
    \draw (0.1,0.5) arc (0:45:0.6);
    \draw (-0.15,0.62) node {$\alpha(z)$};
\draw(-0.5,0.5) node {$*$};
\draw(-0.6,0.5) node[left] {$z$};
  \end{tikzpicture}
  \quad,
\end{equation}
Secondly, if we have a function $f: \mathbb{Z}^2 \rightarrow \mathbb{C}$, and an injection $\iota:\mathbb{Z}^2 \rightarrow \mathbb{C}$, then there is a unique map $F:\iota(\mathbb{Z}^2)\to  \mathbb{C}$ defined by $F(\iota(\vec{r}))=f(\vec{r})$. At the potential risk of confusion we shall denote this map $F$ by the same symbol $f(z)$ and refer to it as the image of $f(\vec{r})$, where $z=\iota(\vec{r})$.

We now wish to define an operator corresponding to the image of the different quasi-local operators $j(\vec r)$ considered in \secref{Chev}. To this end, let us first refine the notation of \secref{Chev}: we use $j_i(\vec r)$ to refer to the quasi-local operator associated with the generator $f_i$, and  $\bj_i(\vec r)$ to refer to that associated with $\bar f_i$. Let us now modify $j_i(z)$ and $\bj_i(z)$ (the images of $j_i(\vec{r})$ and $\bj_i(\vec{r})$ under our embedding into the complex plane) in order to define operators $\phi_i(z)$ and $\bar \phi_i(z)$ as follows\footnote{Here and elsewhere in this paper $i$ denotes $\sqrt{-1}$ when it does not appear as a subscript.}
\begin{align}
  \phi_i (z) &= e^{-i \alpha(z)} j_i(z) \,, \label{eq:pf1} \\
  \bar\phi_i (z) &= e^{+i \alpha(z)} \bj_i(z) \,. \label{eq:pf2}
\end{align}
These operators obey the four-term relations 
\begin{align}
    \delta z_1\, \phi_i(z_1) + \delta z_2\, \phi_i(z_2)
    +  \delta z_3\, \phi_i(z_3) + \delta z_4\, \phi_i(z_4) &= 0 \,,\label{eq:dh1} \\
    \overline{\delta z}_1\, \bar\phi_i(z_1) + \overline{\delta  z}_2\, \bar\phi_i(z_2)
    + \overline{\delta z}_3\,  \bar\phi_i(z_3) + \overline{\delta z}_4\, \bar\phi_i(z_4) &= 0 \,,\label{eq:adh1}
\end{align}
where $\delta z_i$ refers to the complex number associated with the corresponding anti-clockwise plaquette edge in \eqref{fig:embed}, and $\overline{\delta z}_i$ is its complex conjugate. As explained in Section~\ref{sec:CR}, the first of these relations is a discrete integral condition around the embedded plaquette, corresponding to one half of the Cauchy-Riemann relations: it might be referred to as a half-discrete-holomorphicity condition. The second relation is similarly a discrete version of one half of the relations related to discrete-antiholomorphicity. 

Let us now consider the spectral parameter dependence of the various currents. From the representation given by (\ref{eq:Vlambda}) and (\ref{eq:fbarrep}) it is clear that we can express
\begin{align*}
  j_i(z) &= e^{-\lambda(z)} \,\mathfrak{j}_i(z) \,, \\
  \bj_i(z) &= e^{+\lambda(z)} \, \bmj_i (z) \,,
\end{align*}
where the operators $\mj$ and $\bmj$ have no explicit spectral parameter dependence and are of the form
\begin{align*}
\mj_0(z) &=
\sigma^+(z) \ot e^{\eta \sigma_3} \ot  e^{\eta \sigma_3} \ot \cdots \\
\mj_1(z) &=
\sigma^-(z) \ot e^{-\eta \sigma_3} \ot  e^{-\eta \sigma_3} \ot \cdots \\
\bmj_0(z) &=
e^\eta \, \sigma^-(z) \ot e^{\eta \sigma_3} \ot  e^{\eta \sigma_3} \ot \cdots \\
\bmj_1(z) &=
e^{\eta} \, \sigma^+(z) \ot e^{-\eta \sigma_3} \ot  e^{-\eta \sigma_3} \ot \cdots
\end{align*}
Noting the relation \eqref{eq:alphalambda}, it follows that we have
\begin{align}
  \phi_i(z) &= e^{-i s_i\alpha(z)} \, \mj_i(z) \,, \label{eq:pf1} \\
  \bar\phi_i(z) &= e^{+i s_i\alpha(z)} \, \bmj_i(z) \,, \label{eq:pf2}
\end{align}
where all angular information is encoded in the `spin' $s_i$ defined as 
\be s_0 = s_1 = 1+\frac{i \eta}{\pi} \,.\label{eq:spins}\ee
Note that this is exactly the internal spin expected from the winding factors $e^{\pm 2\eta}$ in \eqref{eq:winding}.

\subsection{CFT interpretation}

In the scaling limit, the critical 6V model is described by a compactified free boson action \cite{DSZ87}:
\begin{equation}
  \mathcal{A}[\vphi] = \frac{g}{4\pi} \int d^2r \, [\nabla\vphi(\vec r)]^2 \,,
  \qquad
  \vphi \equiv \vphi + 2\pi \,,
\end{equation}
where $g=1+i\eta/\pi$. By convention, the bosonic field $\vphi$ is the coarse-grained limit of a height variable defined on the dual lattice, with jumps $\pm \pi$ according to the sign of arrows in the 6V configuration. The central charge is $c=1$. The primary operators in this theory are indexed by integer `electric' and `magnetic' charges $(e,m)$, and have conformal dimensions:
\begin{equation}
  \Delta_{e,m} = \frac{(e+mg)^2}{4g} \,,
  \qquad
  \bar\Delta_{e,m} = \frac{(e-mg)^2}{4g} \,.
\end{equation}
More generally, the quasi-local operator inserting a tail of the form $e^{i\alpha \sigma_3} \otimes e^{i\alpha \sigma_3} \otimes \dots$ corresponds to a non-integer electric charge $e \in \alpha/\pi + \mathbb{Z}$. On the other hand, the magnetic charge is always an integer, and a magnetic charge $m$ sitting at point $\vec r$ corresponds to the defect of $\vphi \to \vphi+2\pi m$ of the bosonic field along a closed contour encircling $\vec r$.

The parafermionic operator $\phi_0$, for instance, has charges $e=1+i\eta/\pi=g$ and $m=1$, which indeed gives conformal dimensions $\Delta=g=s_0$ and $\bar\Delta=0$. Similarly, $\phi_1$ has $(e,m)=(-g,-1)$, whereas $\bar\phi_0$ has $(e,m)=(g,-1)$, and $\bar\phi_1$ has $(e,m)=(-g,1)$.

In the case $\eta=0$,  $\phi_0(z)$ and $\phi_1(z)$ are both spin-1 operators and are the $\widehat{\mathfrak{sl}(2)}_1$ lattice operators discussed in \cite{BF91}. In the continuum limit they scale to the corresponding Wess-Zumino-Witten model currents, which are spin-1 Virasoro primaries \cite{MR1424041}. 

\section{Conserved  Currents in the SOS Model}
\label{sec:SOScurrent}

In this section we will examine the behaviour of the currents $j_i, \bj_i$ of Section 4 when we dress the boundary of the vertex model with Baxter intertwiners and use the vertex-face correspondence. In this way we will obtain conserved currents in the SOS model.

\subsection{Dressed Chevalley generators}
\label{sec:dressed}

Let us define the following functions involving Chevalley generators dressed with our Baxter intertwiners:
\be
F_i \trwt a b c \lambda  &=&  \psi^*(a,c|\lambda) f_i \,\psi(a,b|\gl) \,,\label{eq:SOSchev1}\\
\bar F_i \trwt a b c \lambda &=&  \psi^*(a,c|\lambda) \bar f_i \, \psi(a,b|\gl) \,,\\
T_i^{-} \sqwt a b c d \lambda &=&  \psi^*(d,c|\lambda) \, t_i^{- 1} \psi(a,b|\gl) \,,\\
T_i^{+} \sqwt a b c d \lambda &=&  \psi'(d,c|\lambda) \, t_i \,\psi(a,b|\gl) \,.\label{eq:SOSchev4}
\ee
Here, we are viewing $\psi(a,b|\gl)$ as an element of  $V_\gl$, and $\psi^*(a,b|\gl), \psi'(a,b|\gl): V_\gl \to \C$, with the Chevalley generators acting on the representation $V_\gl$ as specified in~\eqref{eq:Vlambda} and \eqref{eq:fbarrep}. Graphically, we have
\begin{align*}
  F_i \trwt a b c \gl , \quad \bar F_i \trwt a b c \gl
  &= \quad \raisebox{-1cm}{\begin{tikzpicture}[scale=0.7]
      \triplaq(0,0,a,b,c);
    \end{tikzpicture}}
  = \quad \raisebox{-1cm}{\begin{tikzpicture}[scale=0.7]
      \draw[gline,fill=fondo] (0,0) node[bblob]{}node[left,black]{$a$} -- (2,1) node[bblob]{}node[right,black]{$b$} -- (2,-1) node[bblob]{}node[right,black]{$c$} -- (0,0);
      \draw(1,0) node{$\lambda$};
      \draw[green!50!black,wavy] (2,0) -- (1.5,0);
      \gcross(1.5,0);
    \end{tikzpicture}}, \\
  T^-_i \sqwt a b c d \lambda
  &= \quad \raisebox{-0.8cm}{\begin{tikzpicture}
      \recmplaq(0,0,a,b,c,d);
    \end{tikzpicture}}
  = \quad \raisebox{-1cm}{\begin{tikzpicture}[scale=0.7]
      \draw[gline,fill=fondo] (-1,1) node[bblob]{}node[left,black]{$a$} -- (1,1) node[bblob]{}node[right,black]{$b$} -- (1,-1) node[bblob]{}node[right,black]{$c$} -- (-1,-1) node[bblob]{}node[left,black]{$d$} -- (-1,1);
      \draw[gline] (-0.5,1) -- (-1,0.5);
      \draw(0,0) node[above] {$\lambda$};
      \draw[green!50!black,wavy] (+1,0) -- (-1,0);
    \end{tikzpicture}}, \\
  T^{+}_i \sqwt a b c d \gl
  &= \quad \raisebox{-0.8cm}{\begin{tikzpicture}
      \recpdashplaq(0,0,a,b,c,d);
    \end{tikzpicture}}
  = \quad \raisebox{-1cm}{\begin{tikzpicture}[scale=0.7]
      \draw[gline,fill=fondo] (-1,1) node[bblob]{}node[left,black]{$a$} -- (1,1) node[bblob]{}node[right,black]{$b$} -- (1,-1) node[bblob]{}node[right,black]{$c$} -- (-1,-1) node[bblob]{}node[left,black]{$d$} -- (-1,1);
      \draw[gline] (-0.5,1) -- (-1,0.5);
      \draw(0,0) node[above] {$\lambda$};
      \draw[green!50!black,wavy] (-1,0) -- (+1,0);
    \end{tikzpicture}},
\end{align*}
where the pictures in the first column are `constructive', in the sense that they follow purely from definitions (\ref{eq:SOSchev1}--\ref{eq:SOSchev4}) and our previous graphical notations. It is also useful to introduce the second column of representations, in which we now view the objects as new SOS weights. The following explicit expressions follow from definitions (\ref{eq:SOSchev1}--\ref{eq:SOSchev4}):
\begin{align}
  F_0 \trwt {a\ }{a \pm 1}{a \pm 1} \gl &=\frac{\pm 1}{2\sinh a\eta} \,,
  \qquad \qquad F_0 \trwt {a\ }{a \mp 1}{a \pm 1} \gl =\frac{\pm e^{\pm a\eta}}{2\sinh a\eta} \,, \label{eq:F1}\\
  F_1 \trwt {a\ }{a \pm 1}{a \pm 1} \gl &=\frac{\mp e^{-2\lambda}}{2\sinh a\eta} \,,
  \qquad \qquad F_1 \trwt {a\ }{a \mp 1}{a \pm 1} \gl = \frac{\mp e^{-2\lambda\mp a\eta}}{2\sinh a\eta} \,,\\
  \bar F_0 \trwt a b c \gl &=e^{2\lambda+\eta} F_1\trwt a b c \gl \,, \qquad
  \bar F_1 \trwt a b c \gl =e^{2\lambda+\eta} F_0\trwt a b c \gl \,, \label{eq:F4}
\end{align}
and
\begin{align} \label{eq:T-weights}
  T^-_i \sqwt a {a \pm 1} {d \pm 1} d \lambda &= \frac{\sinh\frac{(d+a \pm 2\mu_i)\eta}{2}}{\sinh d\eta} \,,
  \qquad T^-_i \sqwt a {a \mp 1} {d \pm 1} d \lambda = \frac{\sinh\frac{(d-a \pm 2\mu_i)\eta}{2}}{\sinh d\eta} \,,
\end{align}
where $\mu_0=1$ and $\mu_1=-1$. The functions $T^+_i$ and $T^-_i$ are related by
\begin{equation} \label{eq:T-symm}
  T_i^+ \sqwt a b c d \lambda = T_i^- \sqwt b a d c \lambda \,.
\end{equation}

\subsection{Elementary moves of SOS tail operators}
\label{sec:tailop}

Let us now list the properties of these SOS versions of Chevalley generators. First of all, they 
obey relations which are the analogue of the intertwining condition \eqref{eq:rdef}. We have
\be
&& \sli_g W\sqwt fgde{\gl_{12}} T_i^- \sqwt abgf{\gl_1} T_i^-\sqwt bcdg{\gl_2}\nn\\
&=&\sli_g  T_i^- \sqwt agef {\gl_2} T_i^- \sqwt gcde {\gl_1} W\sqwt abcg{\gl_{12}} \,,
\label{eq:ybt1}\\[3mm]
&& \sli_g W\sqwt fgde{\gl_{12}} T_i^{+} \sqwt abgf {\gl_1} T_i^{+} \sqwt bcdg {\gl_2}\nn\\
&=&\sli_g  T_i^{+} \sqwt agef {\gl_2} T_i^{+} \sqwt gcde {\gl_1} W\sqwt abgc{\gl_{12}} \,,
\label{eq:ybt2}
\ee
which may be drawn as
\begin{equation}
  \sum_g
  \raisebox{-1.5cm}{\begin{tikzpicture}[scale=0.7]
      \draw[gline,fill=fondo] (0,2)node[left,black]{$f$}--(1,3)node[bblob]{}node[above,black]{$a$}--(3,3)node[above,black]{$b$}--(3,1)node[bblob]{}node[right,black]{$c$}--(2,0)node[below,black]{$d$}--(2,2)--(0,2);
      \draw[gline] (0,2)node[bblob]{}--(0,0)node[bblob]{}node[below,black]{$e$}--(2,0)node[bblob]{};
      \draw[gline] (2,2)node[bblob]{}--(3,3)node[bblob]{};
      \draw (1.7,1.7) node{$g$};
      \draw[gline] (0.5,2)--(0,1.5);
      \draw[gline] (0.75,2.75)--(1.5,3);
      \draw[gline] (2.75,2.75)--(3,2.5);
      \draw[green!50!black,wavy] (2.5,0.5)--(2.5,2.5);
      \draw[green!50!black,wavy] (2.5,2.5)--(0.5,2.5);
    \end{tikzpicture}}
  \quad=\quad \sum_g
  \raisebox{-1.5cm}{\begin{tikzpicture}[scale=0.7]
      \draw[gline,fill=fondo] (0,2)node[bblob]{}node[left,black]{$f$}--(1,3)node[above,black]{$a$}--(1,1)--(3,1)node[black,right]{$c$}--(2,0)node[bblob]{}node[black,below]{$d$}--(0,0)--(0,2);
      \draw[gline] (1,3)node[bblob]{}--(3,3)node[bblob]{}node[black,above]{$b$}--(3,1)node[bblob]{};
      \draw[gline] (0,0)node[bblob]{}node[below,black]{$e$}--(1,1)node[bblob]{};
      \draw (1.4,1.4) node{$g$};
      \draw[gline] (1.5,3)--(1,2.5);
      \draw[gline] (0.75,0.75)--(1.5,1);
      \draw[gline] (0.75,2.75)--(1,2.5);
      \draw[green!50!black,wavy] (2.5,0.5)--(0.5,0.5);
      \draw[green!50!black,wavy] (0.5,0.5)--(0.5,2.5);
    \end{tikzpicture}}
  \quad,\label{fig:SOStailmove1}
\end{equation}
and
\begin{equation}
  \sum_g
  \raisebox{-1.5cm}{\begin{tikzpicture}[scale=0.7]
      \draw[gline,fill=fondo] (0,2)node[left,black]{$f$}--(1,3)node[bblob]{}node[above,black]{$a$}--(3,3)node[above,black]{$b$}--(3,1)node[bblob]{}node[right,black]{$c$}--(2,0)node[below,black]{$d$}--(2,2)--(0,2);
      \draw[gline] (0,2)node[bblob]{}--(0,0)node[bblob]{}node[below,black]{$e$}--(2,0)node[bblob]{};
      \draw[gline] (2,2)node[bblob]{}--(3,3)node[bblob]{};
      \draw (1.7,1.7) node{$g$};
      \draw[gline] (0.5,2)--(0,1.5);
      \draw[gline] (0.75,2.75)--(1.5,3);
      \draw[gline] (2.75,2.75)--(3,2.5);
      \draw[green!50!black,wavy] (0.5,2.5)--(2.5,2.5);
      \draw[green!50!black,wavy] (2.5,2.5)--(2.5,0.5);
    \end{tikzpicture}}
  \quad=\quad \sum_g
  \raisebox{-1.5cm}{\begin{tikzpicture}[scale=0.7]
      \draw[gline,fill=fondo] (0,2)node[bblob]{}node[left,black]{$f$}--(1,3)node[above,black]{$a$}--(1,1)--(3,1)node[black,right]{$c$}--(2,0)node[bblob]{}node[black,below]{$d$}--(0,0)--(0,2);
      \draw[gline] (1,3)node[bblob]{}--(3,3)node[bblob]{}node[black,above]{$b$}--(3,1)node[bblob]{};
      \draw[gline] (0,0)node[bblob]{}node[below,black]{$e$}--(1,1)node[bblob]{};
      \draw (1.4,1.4) node{$g$};
      \draw[gline] (1.5,3)--(1,2.5);
      \draw[gline] (0.75,0.75)--(1.5,1);
      \draw[gline] (0.75,2.75)--(1,2.5);
      \draw[green!50!black,wavy] (0.5,2.5)--(0.5,0.5);
      \draw[green!50!black,wavy] (0.5,0.5)--(2.5,0.5);
    \end{tikzpicture}}
  \quad.\label{fig:SOStailmove2}
\end{equation}
The following inversion relations follow from the corresponding relations \eqref{eq:inv}:
\begin{subequations} \label{eq:sosinv}
  \begin{align}
    \sli_{e} T_i^{+}\sqwt  d e a c \gl   T_i^{-}\sqwt  b a e d \gl
    &= \delta_{bc} \,, \label{eq:sosinv1} \\
    \sli_{e} \frac{\sinh e \eta}{\sinh a \eta} \, T_i^{-} \sqwt d e a c \gl   T_i^{+} \sqwt b a e d \gl
    &= \frac{\sinh d \eta}{\sinh b \eta} \, \delta_{bc} \,, \label{eq:sosinv2} \\
    \sli_{e} T_i^{-} \sqwt e d c a \gl   T_i^{+} \sqwt a b d e \gl
    &= \delta_{bc} \,, \label{eq:sosinv3} \\
    \sli_{e} \frac{\sinh e \eta}{\sinh a \eta} \, T_i^{+}\sqwt  e d c a \gl   T_i^{-}\sqwt  a b d e \gl 
    &= \frac{\sinh d \eta}{\sinh b \eta} \, \delta_{bc} \,. \label{eq:sosinv4}
  \end{align}
\end{subequations}
The diagrammatic realisation of these four relations will prove useful in later discussions. They are respectively 
\begin{subequations}
  \begin{align}
    \sum_e \quad
    \raisebox{-1cm}{\begin{tikzpicture}[scale=0.7]
        \draw[gline,fill=fondo] (0,0)node[bblob]{}node[black,left]{$c$}--(0,2)node[bblob]{}node[black,left]{$b$}--(2,2)node[bblob]{}node[black,right]{$a$}--(2,0)node[bblob]{}node[black,right]{$a$}--(0,0);
        \draw[gline](0,1)node[bblob]{}node[black,left]{$d$}--(2,1)node[bblob]{}node[black,right]{$e$};
        \draw[gline](0,0.6)--(0.4,1);
        \draw[gline](0,1.6)--(0.4,2);
        \draw[green!50!black,wavy](0,0.5)--(2,0.5);
        \draw[green!50!black,wavy](2,0.5)--(3,0.5)--(3,1.5)--(2,1.5);
        \draw[green!50!black,wavy](2,1.5)--(0,1.5);
      \end{tikzpicture}}
    \quad &= \quad \delta_{bc} \,, \\
    \sum_e \frac{\sinh e\eta}{\sinh a\eta} \quad
    \raisebox{-1cm}{\begin{tikzpicture}[scale=0.7]
        \draw[gline,fill=fondo] (0,0)node[bblob]{}node[black,left]{$c$}--(0,2)node[bblob]{}node[black,left]{$b$}--(2,2)node[bblob]{}node[black,right]{$a$}--(2,0)node[bblob]{}node[black,right]{$a$}--(0,0);
        \draw[gline](0,1)node[bblob]{}node[black,left]{$d$}--(2,1)node[bblob]{}node[black,right]{$e$};
        \draw[gline](0,0.6)--(0.4,1);
        \draw[gline](0,1.6)--(0.4,2);
        \draw[green!50!black,wavy](0,1.5)--(2,1.5);
        \draw[green!50!black,wavy](2,1.5)--(3,1.5)--(3,0.5)--(2,0.5);
        \draw[green!50!black,wavy](2,0.5)--(0,0.5);
      \end{tikzpicture}}
    \quad &= \quad \frac{\sinh d\eta}{\sinh b\eta} \, \delta_{bc} \,, \\
    \sum_e \quad
    \raisebox{-1cm}{\begin{tikzpicture}[scale=0.7]
        \draw[gline,fill=fondo] (0,0)node[bblob]{}node[black,left]{$a$}--(0,2)node[bblob]{}node[black,left]{$a$}--(2,2)node[bblob]{}node[black,right]{$b$}--(2,0)node[bblob]{}node[black,right]{$c$}--(0,0);
        \draw[gline](0,1)node[bblob]{}node[black,left]{$e$}--(2,1)node[bblob]{}node[black,right]{$d$};
        \draw[gline](0,0.6)--(0.4,1);
        \draw[gline](0,1.6)--(0.4,2);
        \draw[green!50!black,wavy](2,0.5)--(0,0.5);
        \draw[green!50!black,wavy](0,0.5)--(-1,0.5)--(-1,1.5)--(0,1.5);
        \draw[green!50!black,wavy](0,1.5)--(2,1.5);
      \end{tikzpicture}}
    \quad &= \quad \delta_{bc} \,, \\
    \sum_e \frac{\sinh e\eta}{\sinh a\eta} \quad
    \raisebox{-1cm}{\begin{tikzpicture}[scale=0.7]
        \draw[gline,fill=fondo] (0,0)node[bblob]{}node[black,left]{$a$}--(0,2)node[bblob]{}node[black,left]{$a$}--(2,2)node[bblob]{}node[black,right]{$b$}--(2,0)node[bblob]{}node[black,right]{$c$}--(0,0);
        \draw[gline](0,1)node[bblob]{}node[black,left]{$e$}--(2,1)node[bblob]{}node[black,right]{$d$};
        \draw[gline](0,0.6)--(0.4,1);
        \draw[gline](0,1.6)--(0.4,2);
        \draw[green!50!black,wavy](2,1.5)--(0,1.5);
        \draw[green!50!black,wavy](0,1.5)--(-1,1.5)--(-1,0.5)--(0,0.5);
        \draw[green!50!black,wavy](0,0.5)--(2,0.5);
      \end{tikzpicture}}
    \quad &= \quad \frac{\sinh d\eta}{\sinh b\eta} \, \delta_{bc} \,.
  \end{align}
\end{subequations}

Finally, we have the relations which are the analogues of the defining relations~\eqref{eq:winding} of the original $\uq$ Chevalley generators:
\begin{align}
  \sli_{d,e} \frac{\sinh d \eta}{\sinh a \eta} \,
  T_i^+ \sqwt d e c a \gl F_i \trwt d c e \lambda  T_i^- \sqwt a b c d \gl
  &= e^{2(1-\mu_i)\eta} \, F_i \trwt a b c \gl \,, \label{eq:soscr1} \\
  \sli_{d,e}  \frac{\sinh e \eta}{\sinh b \eta} \,
  T_i^- \sqwt d b c a  \gl F_i \trwt d e b \lambda  T_i^+ \sqwt a b e d \gl
  &= e^{2(\mu_i-1)\eta} \, F_i \trwt a b c \gl \,, \label{eq:soscr2} \\
  \sli_{d,e} \frac{\sinh d \eta}{\sinh a \eta}  \,
  T_i^+ \sqwt d e c a \gl \bar F_i \trwt d c e \lambda  T_i^- \sqwt a b c d \gl
  &=  e^{2(\mu_i-1)\eta} \, \bar F_i \trwt a b c \gl \,, \label{eq:soscr3} \\
  \sli_{d,e}  \frac{\sinh e \eta}{\sinh b \eta}  \,
  T_i^- \sqwt d b c a \gl \bar F_i \trwt d e b \lambda  T_i^+ \sqwt a b e d \gl
  &= e^{2(1-\mu_i)\eta} \, \bar F_i\trwt a b c \gl \,, \label{eq:soscr4}
\end{align}
where we have set $\mu_0=1$ and $\mu_1=-1$ as in \secref{dressed}.
Note the change from $e^{\pm 2\eta}$ to $\{1,e^{\pm 4 \eta}\}$ factors on the right-hand side when compared with \eqref{eq:winding}.  The diagrammatic versions of (\ref{eq:soscr1}--\ref{eq:soscr2}) are:
\begin{align}
  \sum_{d,e} \frac{\sinh d\eta}{\sinh a\eta}
  \quad \raisebox{-1.5cm}{\begin{tikzpicture}[scale=0.7]
      \draw[gline,fill=fondo] (0,-1)node[bblob]{}node[black,left]{$a$}--(0,1)node[bblob]{}node[black,left]{$a$}--(2,2)node[bblob]{}node[black,right]{$b$}--(2,-2)node[bblob]{}node[black,right]{$c$}--(0,-1);
      \draw[gline] (2,1)node[bblob]{}node[black,right]{$c$}--(0,0)node[bblob]{}node[black,left]{$d$}--(2,-1)node[bblob]{}node[black,right]{$e$};
      \draw[gline] (0,0.7)--(0.5,1.25);
      \draw[gline] (0,-0.3)--(0.5,-0.25);
      \draw[green!50!black,wavy] (2,1.5)--(0,0.5);
      \draw[green!50!black,wavy] (0,0.5)--(-1,0.5)--(-1,-0.5)--(0,-0.5);
      \draw[green!50!black,wavy] (0,-0.5)--(2,-1.5);
      \draw[green!50!black,wavy] (2,-1.5)--(3,-1.5)--(3,0)--(1,0);
      \gcross(1,0);
    \end{tikzpicture}}
  \quad &= \quad e^{2(1-\mu_i)\eta}
  \quad \raisebox{-1cm}{\begin{tikzpicture}[scale=0.7]
      \draw[gline,fill=fondo] (0,0) node[bblob]{}node[left,black]{$a$} -- (2,1) node[bblob]{}node[right,black]{$b$} -- (2,-1) node[bblob]{}node[right,black]{$c$} -- (0,0);
      \draw[green!50!black,wavy] (2,0) -- (1.5,0);
      \gcross(1.5,0);
    \end{tikzpicture}} \,, \\
  \sum_{d,e} \frac{\sinh d\eta}{\sinh a\eta}
  \quad \raisebox{-1.5cm}{\begin{tikzpicture}[scale=0.7]
      \draw[gline,fill=fondo] (0,-1)node[bblob]{}node[black,left]{$a$}--(0,1)node[bblob]{}node[black,left]{$a$}--(2,2)node[bblob]{}node[black,right]{$b$}--(2,-2)node[bblob]{}node[black,right]{$c$}--(0,-1);
      \draw[gline] (2,1)node[bblob]{}node[black,right]{$e$}--(0,0)node[bblob]{}node[black,left]{$d$}--(2,-1)node[bblob]{}node[black,right]{$b$};
      \draw[gline] (0,0.7)--(0.5,1.25);
      \draw[gline] (0,-0.3)--(0.5,-0.25);
      \draw[green!50!black,wavy] (0,0.5)--(2,1.5);
      \draw[green!50!black,wavy] (0,-0.5)--(-1,-0.5)--(-1,+0.5)--(0,+0.5);
      \draw[green!50!black,wavy] (2,-1.5)--(0,-0.5);
      \draw[green!50!black,wavy] (2,1.5)--(3,1.5)--(3,0)--(1,0);
      \gcross(1,0);
    \end{tikzpicture}}
  \quad &= \quad e^{2(\mu_i-1)\eta}
  \quad \raisebox{-1cm}{\begin{tikzpicture}[scale=0.7]
      \draw[gline,fill=fondo] (0,0) node[bblob]{}node[left,black]{$a$} -- (2,1) node[bblob]{}node[right,black]{$b$} -- (2,-1) node[bblob]{}node[right,black]{$c$} -- (0,0);
      \draw[green!50!black,wavy] (2,0) -- (1.5,0);
      \gcross(1.5,0);
    \end{tikzpicture}} \,,
\end{align}
and similarly for (\ref{eq:soscr3}--\ref{eq:soscr4}), with inverse phase factors on the right-hand side.

\subsubsection{Generalised SOS tail operators}
It is worth noting that all the elementary moves described so far in \secref{tailop} are also satisfied by a larger class of face weights for tails:
\begin{align}
  \tau_\mu^- \sqwt a b c d \gl &= \psi^*(d,c|\lambda) \mat{e^{+\mu\eta}}{0}{0}{e^{-\mu\eta}} \psi(a,b|\lambda) \,, \\
  \tau_\mu^+ \sqwt a b c d \gl &= \psi'(d,c|\lambda) \mat{e^{-\mu\eta}}{0}{0}{e^{+\mu\eta}} \psi(a,b|\lambda) \,,
\end{align}
for any $\mu \in \mathbb{C}$. The dressed generators $T_0^{\pm}$ and $T_1^{\pm}$ simply correspond to $\mu=1$ and $\mu=-1$ respectively (see \secref{dressed}).

\subsection{Basic four-term relations}

The face weights $F_i$ and $T_i^\pm$ obey four-term relations:

\be  
&& {} \hspace*{-10mm}\sli_g W\sqwt agde{\gl_{12}} F_i \trwt abg{\gl_1} T_i^- \sqwt bcdg{\gl_2}+
W\sqwt{a}{b}{d}{e}{\gl_{12}}  F_i \trwt b c d {\gl_2} \nn \\
&&\hspace*{-10mm} =\sli_g  F_i \trwt a g e  {\gl_2} T_i^- \sqwt g c d e {\gl_1} W\sqwt abcg{\gl_{12}}+
F_i \trwt e c d  {\gl_1}  W\sqwt{a}{b}{c}{e}{\gl_{12}},\label{eq:sosint1}
\\[3mm]
&& {} \hspace*{-10mm}\sli_g W\sqwt agde{\gl_{12}} \bar F_i \trwt a b g  {\gl_1} T_i^- \sqwt b c d g {\gl_2}+
W\sqwt {a}{b}{d}{e}{\gl_{12}}  \bar F_i \trwt b c d {\gl_2} \nn \\
&&\hspace*{-10mm} =\sli_g  \bar F_i \trwt a g e  {\gl_2} T_i^- \sqwt g c d e {\gl_1} W\sqwt{a}{b}{c}{g}{\gl_{12}}+
\bar F_i \trwt e c d  {\gl_1}  W\sqwt{a}{b}{c}{e}{\gl_{12}},\label{eq:sosint2}
\ee
which may be drawn as 
\begin{equation}
  \sum_g \quad
  \raisebox{-1cm}{\begin{tikzpicture}[scale=0.5]
      \draw[gline,fill=fondo] (2,2)--(2,0)node[bblob]{}node[below,black]{$d$}--(3,0)node[right,black]{$c$}node[bblob]{}--(3,2)node[right,black]{$b$}node[bblob]{}--(2,2);
      \draw[gline,fill=fondo] (0,2.5)--(2,2)node[left,black]{$g$}node[bblob]{}--(2,3)node[above,black]{$b$}node[bblob]{}--(0,2.5);
      \draw[gline] (0,2.5)node[left,black]{$a$}node[bblob]{}--(0,0)node[below,black]{$e$}node[bblob]{}--(2,0)node[bblob]{};
      \draw[gline] (0,2)--(0.5,2.375);
      \draw[gline] (2.7,2)--(3,1.7);
      \draw[green!50!black,wavy] (2.5,0)--(2.5,2.5);
      \draw[green!50!black,wavy] (2.5,2.5)--(1,2.5);
      \draw[gline] (0.9,2.6)--(1.1,2.4);
      \draw[gline] (0.9,2.4)--(1.1,2.6);
    \end{tikzpicture}}
  \quad + \quad
  \raisebox{-1cm}{\begin{tikzpicture}[scale=0.5]
      \draw[gline,fill=fondo] (2,0)--(3,0)node[below,black]{$c$}node[bblob]{}--(2.5,2)--(2,0);
      \draw[gline] (2,0)node[below,black]{$d$}node[bblob]{}--(0,0)node[below,black]{$e$}node[bblob]{}--(0,2)node[above,black]{$a$}node[bblob]{}--(2.5,2)node[above,black]{$b$}node[bblob]{};
      \draw[gline] (0,1.5)--(0.5,2);
      \draw[green!50!black,wavy] (2.5,0)--(2.5,1);
      \draw[gline] (2.4,0.9)--(2.6,1.1);
      \draw[gline] (2.4,1.1)--(2.6,0.9);
    \end{tikzpicture}}
  \quad =\sum_g \quad
  \raisebox{-1cm}{\begin{tikzpicture}[scale=0.5]
      \draw[gline,fill=fondo] (0,0)--(2,0)node[right,black]{$c$}--(2,-1)node[below,black]{$d$}node[bblob]{}--(0,-1)node[below,black]{$e$}node[bblob]{}--(0,0);
      \draw[gline,fill=fondo] (-0.5,2)node[above,black]{$a$}--(-1,0)node[left,black]{$e$}node[bblob]{}--(0,0)node[above,black]{$g$}node[bblob]{}--(-0.5,2);
      \draw[gline] (-0.5,2)node[bblob]{}--(2,2)node[above,black]{$b$}node[bblob]{}--(2,0)node[bblob]{};
      \draw[gline] (0,2)--(-0.375,1.625);
      \draw[gline] (0,-0.3)--(0.3,0);
      \draw[green!50!black,wavy] (2,-0.5)--(-0.5,-0.5);
      \draw[green!50!black,wavy] (-0.5,-0.5)--(-0.5,1);
      \draw[gline] (-0.6,0.9)--(-0.4,1.1);
      \draw[gline] (-0.6,1.1)--(-0.4,0.9);
    \end{tikzpicture}}
  \quad + \quad
  \raisebox{-1cm}{\begin{tikzpicture}[scale=0.5]
      \draw[gline,fill=fondo] (0,-0.5)--(2,0)--(2,-1)node[right,black]{$d$}node[bblob]{}--(0,-0.5);
      \draw[gline] (0,-0.5)node[left,black]{$e$}node[bblob]{}--(0,2)node[left,black]{$a$}node[bblob]{}--(2,2)node[right,black]{$b$}node[bblob]{}--(2,0)node[right,black]{$c$}node[bblob]{};
      \draw[gline] (0,1.5)--(0.5,2);
      \draw[green!50!black,wavy] (2,-0.5)--(1,-0.5);
      \draw[gline] (0.9,-0.6)--(1.1,-0.4);
      \draw[gline] (0.9,-0.4)--(1.1,-0.6);
    \end{tikzpicture}}
  \quad.\label{fig:SOS4term}
\end{equation}
The proof of these relations follows from the vertex-face correspondence (\ref{pic:VIRF1}--\ref{pic:VIRF2}), the vertex model intertwining conditions (\ref{eq:tailmove1}--\ref{eq:fint}), and the inversion relations \eqref{eq:inv-a}. The proofs are most easily seen diagrammatically. For example, the proof of \eqref{eq:sosint1} proceeds by the following diagrammatic equalities:
\ben
&&\begin{tikzpicture}
  \draw(-1,0.75) node[] {$\sli_g$};
  \draw[gline] (0,1.5)-- (1.5,1.5) -- (1.5,0);
  \sosplaq(0.5,0.5);
  \draw(0,1.5) node[bblob] {};
  \draw(1.5,1.5) node[bblob] {};
  \draw(1.5,0) node[bblob] {};
  \draw(0,1) node[bblob] {};
  \draw(1,1) node[bblob] {};
  \draw(1,0) node[bblob] {};
  \draw[aline=1] (0.5,1.5) -- (0.5,1);
  \draw[aline=1] (1.5,0.5) -- (1,0.5);
  \draw[blue,wavy] (1.25,0) -- (1.25,1.25) -- (0.5,1.25);
  \scross(0.5,1.25);
  \draw(-0.2,1) node[] {$a$};
  \draw(-0.2,-0.2) node[] {$e$};
  \draw(1,-0.2) node[] {$d$};
  \draw(-0.2,1.7) node[] {$a$};
  \draw(1.7,1.7) node[] {$b$};
  \draw(1.7,-0.2) node[] {$c$};
  \draw(0.7,0.7) node[] {$g$};
\end{tikzpicture}\quad 
\begin{tikzpicture}
  \draw(-1,0.75) node[] {$+\sli_g$};
  \draw[gline] (0,1.5)-- (1.5,1.5) -- (1.5,0);
  \sosplaq(0.5,0.5);
  \draw(0,1.5) node[bblob] {};
  \draw(1.5,1.5) node[bblob] {};
  \draw(1.5,0) node[bblob] {};
  \draw(0,1) node[bblob] {};
  \draw(1,1) node[bblob] {};
  \draw(1,0) node[bblob] {};
  \draw[aline=1] (0.5,1.5) -- (0.5,1);
  \draw[aline=1] (1.5,0.5) -- (1,0.5);
  \draw[blue,wavy] (1.25,-0.2) -- (1.25,0.5);
  \scross(1.25,0.5);
  \draw(-0.2,1) node[] {$a$};
  \draw(-0.2,-0.2) node[] {$e$};
  \draw(1,-0.2) node[] {$d$};
  \draw(-0.2,1.7) node[] {$a$};
  \draw(1.7,1.7) node[] {$b$};
  \draw(1.7,-0.2) node[] {$c$};
  \draw(0.7,0.7) node[] {$g$};
\end{tikzpicture}\quad 
\\&&
\begin{tikzpicture}
  \draw(-1,0.75) node[] {$=$};
  \draw[gline] (0,1.5)-- (1.5,1.5) -- (1.5,0);
  \draw[gline] (0,1) --(0,0) --(1,0);
  \draw(0,1.5) node[bblob] {};
  \draw(1.5,1.5) node[bblob] {};
  \draw(1.5,0) node[bblob] {};
  \draw(0,1) node[bblob] {};
  \draw(1,0) node[bblob] {};
  \draw[aline=1] (0.5,1.5) -- (0.5,0);
  \draw[aline=1] (1.5,0.5) -- (0,0.5);
  \draw[blue,wavy] (1.25,0) -- (1.25,1.25) -- (0.5,1.25);
  \scross(0.5,1.25);
  \draw(-0.2,1) node[] {$a$};
  \draw(-0.2,-0.2) node[] {$e$};
  \draw(1,-0.2) node[] {$d$};
  \draw(-0.2,1.7) node[] {$a$};
  \draw(1.7,1.7) node[] {$b$};
  \draw(1.7,-0.2) node[] {$c$};
\end{tikzpicture}\quad 
\begin{tikzpicture}
  \draw(-1,0.75) node[] {$+$};
  \draw[gline] (0,1.5)-- (1.5,1.5) -- (1.5,0);
  \draw[gline] (0,1) --(0,0) --(1,0);
  \draw(0,1.5) node[bblob] {};
  \draw(1.5,1.5) node[bblob] {};
  \draw(1.5,0) node[bblob] {};
  \draw(0,1) node[bblob] {};
  \draw(1,0) node[bblob] {};
  \draw[aline=1] (0.5,1.5) -- (0.5,0);
  \draw[aline=1] (1.5,0.5) -- (0,0.5);
  \draw[blue,wavy] (1.25,-0.2) -- (1.25,0.5);
  \scross(1.25,0.5);
  \draw(-0.2,1) node[] {$a$};
  \draw(-0.2,-0.2) node[] {$e$};
  \draw(1,-0.2) node[] {$d$};
  \draw(-0.2,1.7) node[] {$a$};
  \draw(1.7,1.7) node[] {$b$};
  \draw(1.7,-0.2) node[] {$c$};
  \draw(4.1,0.75) node[] {using \eqref{pic:VIRF1},};
\end{tikzpicture}
\\&&
\begin{tikzpicture}
  \draw(-1,0.75) node[] {$=$};
  \draw[gline] (0,1.5)-- (0,0) -- (1.5,0);
  \draw[gline] (0.5,1.5) -- (1.5,1.5) -- (1.5,0.5);
  \draw(0,1.5) node[bblob] {};
  \draw(0,0) node[bblob] {};
  \draw(1.5,0) node[bblob] {};
  \draw(.5,1.5) node[bblob] {};
  \draw(1.5,0.5) node[bblob] {};
  \draw[aline=1] (1.5,1) -- (0,1);
  \draw[aline=1] (1,1.5) -- (1,0);
  \draw[blue,wavy](1.5,0.25)  -- (.25,.25) -- (.25,1) ;
  \scross(0.25,1);
  \draw(-0.2,1.7) node[] {$a$};
  \draw(-0.2,-0.2) node[] {$e$};
  \draw(1.7,-0.2) node[] {$d$};
  \draw(0.5,1.7) node[] {$a$};
  \draw(1.7,1.7) node[] {$b$};
  \draw(1.7,0.5) node[] {$c$};
  %
\end{tikzpicture}
\begin{tikzpicture}
  \draw(-1,0.75) node[] {$+$};
  \draw[gline] (0,1.5)-- (0,0) -- (1.5,0);
  \draw[gline] (0.5,1.5) -- (1.5,1.5) -- (1.5,0.5);
  \draw(0,1.5) node[bblob] {};
  \draw(0,0) node[bblob] {};
  \draw(1.5,0) node[bblob] {};
  \draw(.5,1.5) node[bblob] {};
  \draw(1.5,0.5) node[bblob] {};
  \draw[aline=1] (1.5,1) -- (0,1);
  \draw[aline=1] (1,1.5) -- (1,0);
  \draw[blue,wavy](1.7,0.25)  -- (1,.25) ;
  \scross(1,.25);
  \draw(-0.2,1.7) node[] {$a$};
  \draw(-0.2,-0.2) node[] {$e$};
  \draw(1.7,-0.2) node[] {$d$};
  \draw(0.5,1.7) node[] {$a$};
  \draw(1.7,1.7) node[] {$b$};
  \draw(1.7,0.5) node[] {$c$};
  %
  \draw(4.5,0.75) node[] {using \eqref{eq:fint},};
\end{tikzpicture}
\\&&
\begin{tikzpicture}
  \draw(-1,0.75) node[] {$=\sli_g$};
  \draw[gline] (0,1.5)-- (0,0) -- (1.5,0);
  \sosplaq(1,1);
  \draw(0,1.5) node[bblob] {};
  \draw(0,0) node[bblob] {};
  \draw(1.5,0) node[bblob] {};
  \draw(.5,1.5) node[bblob] {};
  \draw(0.5,.5) node[bblob] {};
  \draw(1.5,0.5) node[bblob] {};
  \draw[aline=1] (0.5,1) -- (0,1);
  \draw[aline=1] (1,0.5) -- (1,0);
  \draw[blue,wavy](1.5,0.25)  -- (.25,.25) -- (.25,1) ;
  \scross(0.25,1);
  \draw(-0.2,1.7) node[] {$a$};
  \draw(-0.2,-0.2) node[] {$e$};
  \draw(1.7,-0.2) node[] {$d$};
  \draw(0.5,1.7) node[] {$a$};
  \draw(1.7,1.7) node[] {$b$};
  \draw(1.7,0.5) node[] {$c$};
  \draw(0.7,0.7) node[] {$g$};
\end{tikzpicture}
\begin{tikzpicture}
  \draw(-1,0.75) node[] {$+\sli_g$};
  \draw[gline] (0,1.5)-- (0,0) -- (1.5,0);
  \sosplaq(1,1);
  \draw(0,1.5) node[bblob] {};
  \draw(0,0) node[bblob] {};
  \draw(1.5,0) node[bblob] {};
  \draw(.5,1.5) node[bblob] {};
  \draw(0.5,.5) node[bblob] {};
  \draw(1.5,0.5) node[bblob] {};
  \draw[aline=1] (0.5,1) -- (0,1);
  \draw[aline=1] (1,0.5) -- (1,0);
  \draw[blue,wavy](1.7,0.25)  -- (1,.25);
  \scross(1,.25);
  \draw(-0.2,1.7) node[] {$a$};
  \draw(-0.2,-0.2) node[] {$e$};
  \draw(1.7,-0.2) node[] {$d$};
  \draw(0.5,1.7) node[] {$a$};
  \draw(1.7,1.7) node[] {$b$};
  \draw(1.7,0.5) node[] {$c$};
  \draw(0.7,0.7) node[] {$g$};
  \draw(4.1,0.75) node[] {using \eqref{pic:VIRF2}.};
\end{tikzpicture}
\een
Applying the inversion relation \eqref{eq:inv-a} to the right-most term of the first and last lines of this set of equalities leads to the relation 
\eqref{eq:sosint1}.

\subsection{SOS currents}
\label{sec:JSOS}

Let us now consider a current insertion in our vertex model with a simple tail configuration of the form of \eqref{eq:simpconf} below: 
\begin{equation}\label{eq:simpconf}
  \begin{tikzpicture}[scale=0.8,baseline=2.4cm]
    \foreach\x in {1,2,3,4,5,6} 
    {
      \draw[aline=0.2] (\x,6.5) -- (\x,0.5);
    }
    \foreach\y in {1,2,3,4,5,6}{
      \draw[aline=0.3] (6.5,\y) -- (0.5,\y);
    }
    \draw[blue,wavy=0.4] (5.5,0.5) -- (5.5,4.5) -- (2,4.5);
    \cross(2,4.5);
  \end{tikzpicture}
  \quad.
\end{equation}
That is, the tail
only travels in a up or left direction from the fixed boundary point to the insertion point. Dressing the boundary\footnote{ Note that dressing the bottom right of the boundary with (the dashed) $\psi'$, as opposed to $\psi$, is just a matter of convenience for this particular tail boundary point; it is useful in order that we can use \mref{eq:inv-c} from the lower right-hand corner.} and using the vertex-face correspondence, we find 
\begin{equation*}
  \raisebox{-2cm}{\begin{tikzpicture}[scale=0.7]
    \foreach\x in {1,2,3,4,5,6} 
    {
      \draw[aline=0.2] (\x,6.5) -- (\x,0.5);
      \draw(\x-0.5,0.5) node[bblob] {};
      \draw(\x-0.5,6.5) node[bblob] {};
    }
    \foreach\y in {1,2,3,4,5,6}{
      \draw[aline=0.3] (6.5,\y) -- (0.5,\y);
      \draw(0.5,\y+0.5) node[bblob] {};
      \draw(6.5,\y+0.5) node[bblob] {};
    }
    \draw(6.5,0.5) node[bblob] {};
    \draw[gline] (0.5,0.5) -- (5.5,0.5);
    \draw[gline,dashed] (5.5,0.5) -- (6.5,0.5);
    \draw[gline] (6.5,0.5) -- (6.5,6.5) -- (0.5,6.5) -- (0.5,0.5);
    \draw[blue,wavy=0.4] (5.5,0.7) -- (5.5,4.5) -- (2,4.5);
    \cross(2,4.5);
  \end{tikzpicture}}
  \quad=\quad
  \raisebox{-2cm}{\begin{tikzpicture}[scale=0.7]
    \foreach\x in {1}{
      \foreach\y in {1,2,3,4,5,6}{
        \sosplaq(\x,\y);
      }}
    \foreach\x in {6.5}{
      \foreach\y in {2,3,6}{
        \sosplaq(\x,\y);
      }}
    \foreach\x in {2,3,4}{
      \foreach\y in {1,2,3}{
        \sosplaq(\x,\y);
      }}
    \foreach\x in {2,3,4,5}{
      \foreach\y in {2,3}{
        \sosplaq(\x,\y);
      }}
    \foreach\x in {2,3,4}{
      \sosplaq(\x,6);
    }
    \rwidelsosplaq(5,5);\rwidesosplaq(5,6);
    \lsosplaq(3,4);\lsosplaq(4,4);\lsosplaq(5,4);
    \usosplaq(3,5);\usosplaq(4,5);\usosplaq(6.5,5);
    \tallsosplaq(6.5,4);
    \lslantsosplaq(2,4); \uslantsosplaq(2,5); 
    \rcasosplaq(6.5,1);
    \rcbsosplaq(5,1);
    \draw[blue,wavy=0.4] (5.75,0.75) -- (5.75,4.5) -- (2,4.5);
    \draw[aline=0.98] (3,4.8) -- (3,4.2);
    \draw[aline=0.98] (4,4.8) -- (4,4.2);
    \draw[aline=0.98] (5,4.8) -- (5,4.2);
    \draw (2,4.65) -- (2,4.35);
    \draw[aline=1] (5.9,1) -- (5.6,1);
    \draw[aline=0.98] (6.0,2) -- (5.5,2);
    \draw[aline=0.98] (6.0,3) -- (5.5,3);
    \draw[aline=0.98] (6.0,3.8) -- (5.5,3.8);
    \cross(2,4.5);
  \end{tikzpicture}}
  = \quad
  \raisebox{-2cm}{\includegraphics[scale=0.55]{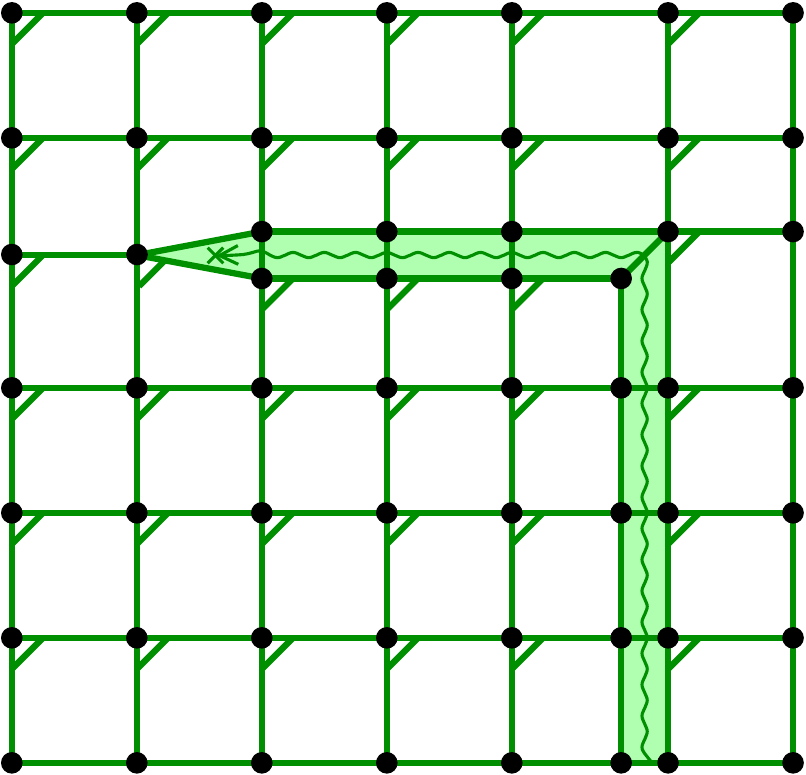}}
\end{equation*}
where all internal height variables are summed over (and the boundary heights either side of the insertion point in the final picture are identified). We see that in this example the dressed quasi-local current in the SOS model is made up from a local insertion of $F_i$ or $\bar F_i$ and a tail consisting of just $T_i^-$ insertions.

Now suppose we consider general tail configurations corresponding to the $f_i$ or $\bar f_i$ currents in our vertex model: these may contain right-going and  down-going tails and may have non-trivial winding. Let us define embedded quasi-local SOS currents
$J_i(z)$ and $\bar J_i(z)$ as consisting of the following (note that after embedding into the complex plane, our SOS lattice squares will appear as tilted rhombi as explained in Section \ref{sec:para-6v}. For simplicity however, we continue to draw them and describe their orientation in terms of the squares that are their preimage):
\begin{itemize}
\item insertion of the operator $F_i$ or $\bar F_i$ at embedded lattice point $z$,
\item insertion of left- and up-going portions of tails consisting of
  $T_i^{-} = \raisebox{-0.5cm}{\begin{tikzpicture}[scale=0.5]
      \draw[gline,fill=fondo] (-1,1) node[bblob]{} -- (1,1) node[bblob]{} -- (1,-1) node[bblob]{} -- (-1,-1) node[bblob]{} -- (-1,1);
      \draw[gline] (-0.5,1) -- (-1,0.5);
      \draw[green!50!black,wavy] (+1,0) -- (-1,0);
    \end{tikzpicture}}$
  as above,
\item insertion of right- and down-going portions of tails consisting  of 
  $T_i^{+}= \raisebox{-0.5cm}{\begin{tikzpicture}[scale=0.5]
      \draw[gline,fill=fondo] (-1,1) node[bblob]{} -- (1,1) node[bblob]{} -- (1,-1) node[bblob]{} -- (-1,-1) node[bblob]{} -- (-1,1);
      \draw[gline] (-0.5,1) -- (-1,0.5);
      \draw[green!50!black,wavy] (-1,0) -- (+1,0);
    \end{tikzpicture}}$,
\item insertion of the additional corner factors:
  \begin{align*}
    &\raisebox{-5mm}{\begin{tikzpicture}[scale=0.5]
      \draw[gline,fill=fondo] (0,3)node[bblob]{}--(3,3)--(3,0)node[bblob]{}--(2,0)node[bblob]{}--(2,2)--(0,2)node[bblob]{}--(0,3);
      \draw[gline] (2,2)node[bblob]{}--(3,3)node[bblob]{};
      \draw[green!50!black,wavy] (2.5,0) -- (2.5,2.5);
      \draw[green!50!black,wavy] (2.5,2.5) -- (0,2.5);
    \end{tikzpicture}}
    \quad,\quad
    \raisebox{-5mm}{\begin{tikzpicture}[scale=0.5]
      \draw[gline,fill=fondo] (0,3)node[bblob]{}--(3,3)--(3,0)node[bblob]{}--(2,0)node[bblob]{}--(2,2)--(0,2)node[bblob]{}--(0,3);
      \draw[gline] (2,2)node[bblob]{}--(3,3)node[bblob]{};
      \draw[green!50!black,wavy] (0,2.5) -- (2.5,2.5);
      \draw[green!50!black,wavy] (2.5,2.5) -- (2.5,0);
    \end{tikzpicture}}
    \quad,\quad
    \raisebox{-5mm}{\begin{tikzpicture}[scale=0.5]
      \draw[gline,fill=fondo] (0,0)node[bblob]{}--(1,0)node[bblob]{}--(1,2)--(3,2)node[bblob]{}--(3,3)node[bblob]{}--(0,3)--(0,0);
      \draw[gline] (0,3)node[bblob]{}node[above,black]{$a$}--(1,2)node[bblob]{}node[right,black]{$b$};
      \draw[green!50!black,wavy] (3,2.5)--(0.5,2.5);
      \draw[green!50!black,wavy] (0.5,2.5)--(0.5,0);
    \end{tikzpicture}}
    \times \, \frac{\sinh b\eta}{\sinh  a\eta }
    \quad,\quad
    \raisebox{-5mm}{\begin{tikzpicture}[scale=0.5]
      \draw[gline,fill=fondo] (0,0)node[bblob]{}--(1,0)node[bblob]{}--(1,2)--(3,2)node[bblob]{}--(3,3)node[bblob]{}--(0,3)--(0,0);
      \draw[gline] (0,3)node[bblob]{}--(1,2)node[bblob]{};
      \draw[green!50!black,wavy] (0.5,0)--(0.5,2.5);
      \draw[green!50!black,wavy] (0.5,2.5)--(3,2.5);
    \end{tikzpicture}} \quad, \\
    & \raisebox{-5mm}{\begin{tikzpicture}[scale=0.5]
      \draw[gline,fill=fondo] (0,0)--(3,0)node[bblob]{}--(3,1)node[bblob]{}--(1,1)--(1,3)node[bblob]{}--(0,3)node[bblob]{}--(0,0);
      \draw[gline] (0,0)node[bblob]{}--(1,1)node[bblob]{};
      \draw[green!50!black,wavy] (3,0.5)--(0.5,0.5);
      \draw[green!50!black,wavy] (0.5,0.5)--(0.5,3);
    \end{tikzpicture}}
    \quad,\quad
    \raisebox{-5mm}{\begin{tikzpicture}[scale=0.5]
      \draw[gline,fill=fondo] (0,0)--(3,0)node[bblob]{}--(3,1)node[bblob]{}--(1,1)--(1,3)node[bblob]{}--(0,3)node[bblob]{}--(0,0);
      \draw[gline] (0,0)node[bblob]{}--(1,1)node[bblob]{};
      \draw[green!50!black,wavy] (0.5,3)--(0.5,0.5);
      \draw[green!50!black,wavy] (0.5,0.5)--(3,0.5);
    \end{tikzpicture}}
    \quad,\quad
    \raisebox{-5mm}{\begin{tikzpicture}[scale=0.5]
      \draw[gline,fill=fondo] (0,0)node[bblob]{}--(3,0)--(3,3)node[bblob]{}--(2,3)node[bblob]{}--(2,1)--(0,1)node[bblob]{}--(0,0);
      \draw[gline] (3,0)node[bblob]{}node[right,black]{$a$}--(2,1)node[bblob]{}node[left,black]{$b$};
      \draw[green!50!black,wavy] (2.5,3)--(2.5,0.5);
      \draw[green!50!black,wavy] (2.5,0.5)--(0,0.5);
    \end{tikzpicture}}
    \times \, \frac{\sinh b\eta}{\sinh  a\eta }
    \quad,\quad
    \raisebox{-5mm}{\begin{tikzpicture}[scale=0.5]
      \draw[gline,fill=fondo] (0,0)node[bblob]{}--(3,0)--(3,3)node[bblob]{}--(2,3)node[bblob]{}--(2,1)--(0,1)node[bblob]{}--(0,0);
      \draw[gline] (3,0)node[bblob]{}--(2,1)node[bblob]{};
      \draw[green!50!black,wavy] (0,0.5)--(2.5,0.5);
      \draw[green!50!black,wavy] (2.5,0.5)--(2.5,3);
    \end{tikzpicture}} \quad.
  \end{align*}
\end{itemize}
That is, all of the corners are formed from the appropriate pairs of $T_i^{\pm}$ with the additional requirement that the two corners involving left $\rightarrow$ down, or down $\rightarrow$ left orientations of the tail acquire the extra factor $\frac{\sinh b\eta}{\sinh  a\eta }$ indicated depending on the height variables on either side of the corner. 

With this definition of SOS currents, we then have the following equivalence under the vertex correspondence
\begin{equation} \label{eq:equiv}
  j_i(z) \sim e^{2M \mu_i \eta} J_i(z), \quad j_i(z) \sim e^{-2M \mu_i \eta} \bar J_i(z) \,, \quad  i\in\{0,1\} \,, \qquad
\end{equation}
where $M$ is defined to be the anti-clockwise winding number of the tail, and $\mu_0=1, \mu_1=-1$. This equivalence is compatible with the Yang-Baxter relations (\ref{eq:ybt1}--\ref{eq:ybt2}), the inversion inversion relations (\ref{eq:sosinv1}--\ref{eq:sosinv4}), and the commutation relations (\ref{eq:soscr1}--\ref{eq:soscr4}), in the following sense: we can either first move the tail using the corresponding relations for the vertex model as discussed in Section~\ref{sec:6VSOS}, and then apply the vertex-face correspondence, or we can first apply the vertex-face correspondence and then manipulate the SOS tail. The result is the same in both cases.

It also follows from the intertwining relations (\ref{eq:sosint1}--\ref{eq:sosint2}) that $J_i(z)$, $\bar J_i(z)$ obey the following four-term current-conservation relation around an SOS plaquette of the form \eqref{eq:embed}:
\begin{equation} \label{eq:SOScurrentcons}
J(z_1) - J(z_2) - J(z_3) +J(z_4) =0 \,.
\end{equation}
As always, these relations are to be interpreted as expectation values in the embedded SOS model.

\subsection{Parafermionic operators in the SOS Model}
\label{sec:para-SOS}

We may now proceed as in the vertex model and define 

\begin{equation} \label{eq:Phi}
  \begin{aligned}
    \Phi_i (z) &= e^{-i \alpha(z)} \, J_i(z) \,, \\
    \bar\Phi_i (z) &= e^{+i \alpha(z)} \, \bar J_i(z) \,,
  \end{aligned}
\end{equation}
where $\alpha(z)$ has the same definition as in \secref{para-6v}.
With these definitions, the SOS current conservation relation \eqref{eq:SOScurrentcons} becomes
\be
  \delta z_1 \,\Phi_i(z_1) + \delta z_2 \,\Phi_i(z_2)
  + \delta z_3 \,\Phi_i(z_3) + \delta z_4 \,\Phi_i(z_4) &= 0 \,, \label{eq:SOSdh}\\
  \overline{\delta z}_1 \,\bar\Phi_i(z_1) + \overline{\delta z}_2 \,\bar\Phi_i(z_2)
  + \overline{\delta z}_3 \,\bar\Phi_i(z_3) + \overline{\delta z}_4 \,\bar\Phi_i(z_4) &= 0 \,.\label{eq:SOSadh}
\ee
As with vertex models we can rewrite these expressions in a form where all the spectral-parameter dependence of $F_i$ and $\bar F_i$ of equations (\ref{eq:F1}--\ref{eq:F4}) is taken into an exponential prefactor. Explicitly, we have
\begin{align}
\Phi_i(z) &= e^{-i s_i \alpha(z)} \, {\cal J}_i(z) \,, \\
\bar\Phi_i(z) &= e^{+i s_i \alpha(z)} \, \bar{\cal J}_i(z) \,,
\end{align}
where
\begin{equation}
  s_0=1 \,, \qquad s_1=1 +\frac{2i\eta}{\pi} \,,\label{eq:SOSspins}
\end{equation}
and
$$
{\cal J}_i(z)= e^{-\frac{2\eta \alpha(z)\delta_{i,1} }{\pi}} J_i(z),
\qquad
\bar{\cal J}_i(z)= e^{\frac{2\eta \alpha(z) \delta_{i,1} }{\pi}} J_i(z)
$$
are spectral-parameter independent operators.

\subsection{Heights adjacent to the tail}
\label{sec:heights}

The face weights~\eqref{eq:T-weights} have the properties:
\begin{equation} \label{eq:Tvanish}
  T_0^- \sqwt {b\pm 1} b b {b \mp 1} \gl = 0 \,,
  \qquad
  T_1^- \sqwt {a \pm 1} {a \pm 2} {a \mp 2} {a \mp 1} \gl = 0 \,.
\end{equation}
Let us first discuss the consequences for the currents $J_0(z)$ and $\bar J_0(z)$. We shall reason on a simple tail configuration\footnote{Generic paths may include $T_i^+$ faces, but the reasoning remains valid, due to the symmetry property \eqref{eq:T-symm}.} made only of $T_0^-$'s as in~\eqref{eq:simpconf}.
If at any point $z'$ on the path of the tail, the height variables across the path are equal, then this is also the case for all pairs of adjacent height variables, along the portion of the path joining $z'$ to the insertion point $z$. Since the height variables are indeed equal on either side of the boundary insertion point, it follows that the tail is only made of faces of the form:
\begin{equation}
  T_0^- \sqwt a b b a \gl = \frac{\sinh b\eta}{\sinh a\eta} \,.
\end{equation}
Note that the product of all these factors simplifies, so $J_0$ and $\bar J_0$ finally take a local form in terms of the height variables at the insertion point of $F_0$ or $\bar F_0$:
\begin{equation} \label{eq:J0-local}
  J_0 \propto \frac{F_0 \trwt a b b \gl}{\sinh b\eta} \,,
\end{equation}
and similarly for $\bar J_0$.
This locality property is consistent with the fact that $\Phi_0$ and $\bar \Phi_0$ have integer internal spin.

For the SOS currents $J_1$ and $\bar J_1$, a similar property propagates along the path, but in the opposite direction, i.e., from the insertion point of $F_1$ or $\bar F_1$ to the boundary. The only allowed faces along the path are of the form $T_1^- \sqwt a b c d \gl$, where $|a-b|=1$, while $|a-d|$ and $|b-c|$ can only take the values $\{-2,0,+2\}$.

\subsection{The RSOS case}
\label{sec:RSOS}

In this section, we briefly discuss the compatibility of the vertex-face correspondence with RSOS restriction of the face weights. We consider the case $\eta=i\pi/(p+1)$ and integer $p \geq 3$. In this case, as is well known \cite{ABF84}, if we set the reference point $x_0$ [see \secref{VFC}] to zero, we can consistently restrict the SOS height variables to lie in the interval $\{ 1, 2 \dots p \}$; the reason is that $W\sqwt{a}{a+1}{a}{a-1}{\lambda}$ in \eqref{eq:W} has a simple zero when $a=p$.

Now consider the vertex-face correspondence \eqref{eq:VIRF2}, represented graphically by \eqref{pic:VIRF2}. In the situation of interest, we have
  \be
\begin{tikzpicture}[scale=0.3,baseline=0] 
    \draw[gline] (-2,2) node[bblob]{} -- (-2,-2) node[bblob]{} -- (2,-2) node[bblob]{};
    \draw[aline=0.25] (0,2) node[above]{$\lambda_1$} -- (0,-2);
    \draw[aline=0.25] (2,0)  node[right]{$\lambda_2$} -- (-2,0);
    \draw(-2,2) node[above]{$p$};
    \draw(-2,-2) node[left]{$p-1$};
    \draw(2,-2) node[right]{$p$};
    \node at (8,-0.5){$\ds = $};
    \def\x{14};
    \draw(\x-2,+2) node[above] {$p$};
    \draw(\x+2,+2) node[above] { $p-1$};
    \draw(\x+2,-2) node[below] {$p$};
    \draw(\x-2,-2) node[below] {$p-1$};
    \draw(\x+0,-1) node[above] {$\lambda_{12}$};
    \draw[gline] (\x-2,2) node[bblob] {} -- (\x+2,2) node[bblob] {} -- (\x+2,-2) node[bblob] {} -- (\x-2,-2) node[bblob] {} -- (\x-2,2);
    \draw[gline] (\x-1,2) -- (\x-2,1);
    \draw[aline=0.5] (\x+0,4) node[above]{$\lambda_1$} -- (\x+0,2);
    \draw[aline=0.5] (\x+4,0) node[right]{$\lambda_2$} -- (\x+2,0);
  \end{tikzpicture}
  \begin{tikzpicture}[scale=0.3,baseline=0] 
       \node at (8,-0.5){$\ds +$};
    \def\x{14};
    \draw(\x-2,+2) node[above] {$p$};
    \draw(\x+2,+2) node[above] { $p+1$};
    \draw(\x+2,-2) node[below] {$p$};
    \draw(\x-2,-2) node[below] {$p-1$};
    \draw(\x+0,-1) node[above] {$\lambda_{12}$};
    \draw[gline] (\x-2,2) node[bblob] {} -- (\x+2,2) node[bblob] {} -- (\x+2,-2) node[bblob] {} -- (\x-2,-2) node[bblob] {} -- (\x-2,2);
    \draw[gline] (\x-1,2) -- (\x-2,1);
    \draw[aline=0.5] (\x+0,4) node[above]{$\lambda_1$} -- (\x+0,2);
    \draw[aline=0.5] (\x+4,0) node[right]{$\lambda_2$} -- (\x+2,0);
    \node at (22,-0.5){$.$}; \end{tikzpicture}
\ee
Here, both terms on the right-hand-side contribute because the zero in the right-most face weight is cancelled by a pole in the intertwiner $\psi^*(p+1,p|\lambda)$. Nevertheless, if we dress the boundary of the vertex model with intertwiners with heights in $\{1,2,\cdots,p\}$ then it turns out that following the procedure of Section 3.2.2 produces a RSOS partition function with summed over heights purely in the range $\{1,2,\cdots,p\}$ -- the reason being that there are further zeros appearing from other products of intertwiners.

However, if we start with a vertex model with current insertions,  then it is a simple exercise to check that applying the vertex-face correspondence, as in \secref{JSOS}, produces non-zero weight contributions from configurations with heights outside of $\{1,2,\cdots,p\}$ adjacent to the tail. Thus our current insertions are not compatible with the RSOS restriction, and there is no simple way of understanding the role of our currents in the CFTs associated with RSOS models.

\section{The Cyclic Solid-On-Solid model}
\label{sec:CSOS}

Throughout this section, we consider the case when $\eta$ is of the form $\eta=i\pi (p-p')/p$, where $p$ and $p'$ are coprime integers, and $p'<p$.

\subsection{The Cyclic SOS Lattice model}

Given the two coprime integers $p$ and $p'$, we define the integers $\ell$ and $n$ as follows:
\begin{equation} \label{eq:ldef}
  \ell = \begin{cases}
    \frac{p-p'}{2} & \text{if $(p-p')$ is even,} \\
    p-p' & \text{if $(p-p')$ is odd,}
  \end{cases}
  \qquad
  n = \begin{cases}
    p & \text{if $(p-p')$ is even,} \\
    2p & \text{if $(p-p')$ is odd.}
  \end{cases}
\end{equation}
With this definition, $\ell$ and $n$ are also coprime, and we have
\begin{equation}
  \eta = \frac{i\pi(p-p')}{p} = \frac{2i\pi \ell}{n} \,.
\end{equation}
Then, under a simultaneous shift of all heights, the expressions (\ref{eq:psi}--\ref{eq:psid}) for Baxter intertwiners $\psi(a,b|\lambda)$ and $\psi^*(a,b|\lambda)$ become $n$-periodic (resp. antiperiodic) for even $\ell$ (resp. odd $\ell$), and the face weights $W \sqwt abcd \lambda$ \eqref{eq:W} become $n$-periodic. Hence, one may easily define the analogous objects for heights living in $x_0+\mathbb{Z}/n\mathbb{Z} \equiv x_0+\{0,\dots,n-1\}$. For instance, the definition~\eqref{eq:psi} of $\psi(a,a+1)$ is kept for $a \in x_0+\{0,\dots,n-2\}$, and completed by:
\begin{equation*}
  \def\arraystretch{1.8}
  \psi(x_0+(n-1),x_0|\lambda) \equiv    \psi(x_0-1,x_0|\lambda)  = \left[ \begin{array}{c} \exp\left( \frac{-\lambda +(x_0-1)\eta}{2} \right) \\ \exp\left( \frac{+\lambda-(x_0-1)\eta}{2}\right) \end{array} \right] \,,
\end{equation*}
and similarly for $\psi(a,a-1|\lambda)$, $\psi^*(a,b|\lambda)$ and $W \sqwt abcd\lambda$.
In this way, the intertwining relations (\ref{eq:VIRF1}--\ref{eq:VIRF2}) and the YBE relations \eqref{eq:YBE-SOS} are transported from the unrestricted SOS model to the cyclic Solid-On-Solid (CSOS) model, i.e., the SOS model with heights living in $x_0+\mathbb{Z}/n\mathbb{Z}$ and critical weights given by \eqref{eq:W}:
\begin{equation*}
  \begin{aligned}
    W\sqwt{a}{a\pm 1}{a\pm 2}{a\pm 1}{\lambda} &= \sinh(\lambda+\eta),\\
    W\sqwt{a}{a\pm 1}{a}{a\mp 1}{\lambda} &= \frac{\sinh\lambda\ \sinh[(a\pm 1)\eta]}{\sinh(a\eta)},\\
    W\sqwt{a}{a\pm 1}{a}{a\pm 1}{\lambda} &= \frac{\sinh\eta\ \sinh(a\eta\mp \lambda)}{\sinh(a\eta)} .
  \end{aligned}
\end{equation*}
 Elliptic CSOS models were introduced and studied in \cite{KunibaYajima87,PearceSeaton89,KimPearce89}, and their correlation functions were computed more recently in \cite{MR3216385,MR3077806}. Note that in early studies of the CSOS model (see \cite{KunibaYajima87,PearceSeaton89,KimPearce89}), a different limit of the elliptic Boltzmann weights was used, leading to trigonometric weights $W \sqwt abcd\lambda$ depending only on differences $(a-b), (b-c), (c-d)$. Here in contrast, we are dealing with {\it dynamical} trigonometric weights, i.e., the function $W \sqwt abcd\lambda$ cannot be expressed solely in terms of $(a-b), (b-c), (c-d)$.

Let us recall the relation to the Temperley-Lieb (TL) loop model \cite{Pasquier87}. The face weights \eqref{eq:W} are of the form 
\begin{equation}
  W \sqwt abcd\lambda = \sinh(\lambda+\eta) \, \delta_{bd} + \sinh \lambda \, E \left(\begin{array}{cc}a&b \\ d&c \end{array}\right) \,,
\end{equation}
where 
\be
E\left(\begin{array}{cc}a&a+\varepsilon_1 \\ a+\varepsilon_2&b \end{array} \right)=
-\ep_1\ep_2 \,\delta_{a,b} \frac{\sinh[(a+\ep_1)\eta]}{\sinh[a\eta]},\quad \ep_1,\ep_2\in\{+,-\} \,.
\ee
The matrix $E\left(\begin{array}{cc}a&b \\ d&c \end{array}\right)$ gives a representation of the generator of the Temperley-Lieb (TL) algebra with loop weight $-2\cosh\eta$.

\subsection{Identification of parafermionic operators in the scaling limit CFT}

For appropriate boundary conditions, the partition function of the CSOS and TL loop models are equal, and hence their central charges too. For a loop weight $-2\cos \frac{\pi(p-p')}{p}$, Coulomb gas arguments have led to the identification of the TL model central charge as \cite{MR751711,MR675241}
\begin{equation} \label{eq:cc}
  c = 1 - \frac{6(p-p')^2}{pp'} \,.
\end{equation}
A generic method to determine the operator content of SOS models was proposed in \cite{Pasquier87,Pasquier-op-content87}, and was also applied to the `non-dynamical' critical CSOS model in~\cite{KimPearce89}. Up to our knowledge, this exercise has not been done for the CSOS model with dynamical weights~\eqref{eq:W} that we consider here, and so we have performed it in Appendix~\ref{appendix}. To summarise, the primary operators are combinations of local height probabilities and height defects, with conformal dimensions
\begin{equation}
  \begin{cases}
    h_{em} =  \frac{(ep-mp')^2 - (p-p')^2}{4pp'} \,, \\
    \\
    \bar h_{em} =  \frac{(ep+mp')^2 - (p-p')^2}{4pp'} \,,
  \end{cases}
  \qquad e \in \mathbb{Z}/n \,, \quad m \in n\mathbb{Z} \,.
\label{eq:param}\end{equation}

Our parafermionic operators $\Phi_0$ and $\bar\Phi_0$   are local (as shown in \secref{heights}) and involve two neighbouring heights $(a,b)$. The construction in \secref{para-SOS} indicates that $\Phi_0$ (resp. $\bar\Phi_0$) can be identified with a holomorphic (resp. anti-holomorphic) field of scaling dimension $h=1$ in the continuum limit. This corresponds to the dimension of screening charges, as opposed to conformal primary fields, in the Coulomb Gas picture.

The parafermionic operator $\Phi_1$ is quasi-local, and from \secref{para-SOS} it can be identified with a holomorphic field of scaling dimension $h=s_1=1+2i\eta/\pi$ (and similarly for  $\bar\Phi_1$ - which may be identified with an antiholomorphic operator). Using the parametrization \eqref{eq:param}, we find that $s_1=h_{13}$, and hence identify $\Phi_1$ with the holomorphic part of the degenerate operator $\Phi_{13}$.

\section{Conclusions}
\label{sec:conclusion}
In this paper, we have used the quantum-group current formalism of \cite{BF91} in order to construct quasi-local operators in the 6V and trigonometric SOS models. For the 6V case, we have described a systematic procedure that starts from the quantum-group conserved currents and arrives at lattice parafermionic operators \eqref{eq:pf1} and \eqref{eq:pf2} that (i) obey the discrete integral conditions \eqref{eq:dh1} and \eqref{eq:adh1} around a rhombus in the complex plane, and (ii) have spins given by \eqref{eq:spins}. These spins appear naturally in \eqref{eq:pf1} and \eqref{eq:pf2} and express the dependence of the operators on the embedding angle shown in \eqref{fig:embed2}. The identification with CFT primary fields is then automatic and is described in Section 4.6. 

For the trigonometric SOS case, we have defined currents of the SOS model as the image of those in the 6V model under the vertex-face correspondence. The resulting quasi-local SOS operators are shown to obey analogous properties to those of the 6V model: for example, the SOS tail can be moved by \eqref{fig:SOStailmove1} and \eqref{fig:SOStailmove2}, and the currents obey the four-term relation \eqref{fig:SOS4term}. Following a very similar procedure to the 6V case, we obtain SOS parafermionic operators obeying the discrete integral conditions \eqref{eq:SOSdh} and \eqref{eq:SOSadh} with spins given by \eqref{eq:SOSspins}.

In considering the CFT interpretation of our SOS parafermionic operators, we have concentrated on the CSOS case corresponding to the choice of anisotropy parameter 
$\eta= i\pi(p-p')/p$, with $p'<p$ coprime. The CSOS case has the twin advantages  that it has a relatively well understood CFT limit, and that our quasi-local operators are compatible with the cyclic 
restriction of height variables (unlike the case for RSOS models as we point out in Section \ref{sec:RSOS}). We show that our operator  $\Phi_0$  actually becomes local in this CSOS case, and we identify it with a CFT screening operator. The operator $\Phi_1$ is still genuinely quasi-local in the CSOS case and is identified  with the chiral $\Phi_{13}$ operator.

There are various questions that are still under investigation by the authors. One problem is to reinterpret the discrete integral conditions away from the critical line by using perturbed conformal field theory, and to identify perturbing fields. Such an analysis was carried out successfully for the chiral Potts model in \cite{IkhWes16}; however, the situation for the 6V model close to $\eta=0$ is more subtle due to the marginal nature of the perturbation associated to the parameter $\eta$. Finally, the important question of whether our approach can be extended in certain models in order to produce fully discretely-holomorphic operators, whose existence would pave the way for rigorous results on the scaling limit, remains open. 

\subsection*{Acknowledgements}
We would like to thank Beno\^{\i}t Estienne, Nikolai Kitanine, V\'eronique Terras and Paul Zinn-Justin for sharing their useful insights and advice. We also acknowledge the support and hospitality of the Galileo Galilei Institute for Theoretical Physics, and the organisers of the Statistical Mechanics, Integrability and Combinatorics workshop at which this work was begun. RAW would like to thank the Laboratoire de Physique Th\'eorique et Hautes Energies (Universit\'e Pierre et Marie Curie/CNRS) for providing warm hospitality and funding during several visits to Paris.


\newpage
\appendix

\section{Operator content in the Scaling Limit of the CSOS Model}
\label{appendix}

\subsection{Local height probabilities}

In this appendix, we sketch the analysis of local operators in the critical CSOS model, using the approach of \cite{Pasquier-op-content87}. The CSOS model with heights in $x_0+\mathbb{Z}/n\mathbb{Z}$ and weights \eqref{eq:W} corresponds to the adjacency matrix
\begin{equation}
  A_{ab} = \begin{cases}
    1 & \text{if } |a-b| \equiv 1 \mod n \,, \\
    0 & \text{otherwise.}
  \end{cases}
\end{equation}
We consider $n$ orthonormal eigenvectors of $A$, which we denote by
\be
    \{ S^{(1)},\cdots,S^{((n-1)/2)}, T^{(0)},T^{(1)},\cdots T^{((n-1)/2)} \}\quad     & \text{for odd $n$,} \\
   \{ S^{(1)},\cdots,S^{(n/2-1)}, T^{(0)},T^{(1)},\cdots T^{(n/2)} \}\quad     & \text{for even $n$.}
 \ee
Components are given by
\begin{equation}
  S^{(j)}_a = N_j \, \sin\left(\frac{2\pi a j}{n}\right) \,,
  \qquad T^{(j)}_a = N_j \, \cos\left(\frac{2\pi a j}{n}\right) \,, \quad a\in\{0,1,\cdots,n-1\}\,,
\end{equation}
where $N_j=\sqrt{1/n}$ if $j=0$ or $j=n/2$, and $N_j=\sqrt{2/n}$ otherwise. 
Let us define (for the appropriate range of $j$ for $n$ odd or even)
\begin{equation}
  \vphi^{(j)}_a = \frac{S^{(j)}_a}{S^{(\ell)}_a} \,,
  \qquad \psi^{(j)}_a = \frac{T^{(j)}_a}{S^{(\ell)}_a} \,,
\end{equation}
where $\ell$ is defined by \eqref{eq:ldef}.
Local operators, which we denote by the same notation $\vphi^{(j)}_a$ and $\psi^{(j)}_a$, have the following effect in the TL graphical expansion~\cite{Pasquier-op-content87}: any loop enclosing a single operator $\vphi^{(j)}_a$ or $\psi^{(j)}_a$ is assigned a weight $-2\cos \frac{2\pi j}{n}$ instead of $-2\cos \frac{2\pi \ell}{n}=-2\cosh \eta$. The difference between $\vphi^{(j)}_a$ or $\psi^{(j)}_a$ appears when a loop encloses a product of operators, which is important for multi-point correlation functions. We shall not discuss this further, since the two-point functions are enough for the determination of critical exponents.

Using standard Coulomb gas arguments for the loop model \cite{MR751711}, one can identify $\vphi^{(j)}_a$ or $\psi^{(j)}_a$ as an electric charge $e=2j/n+k$, where $k \in \mathbb{Z}$, and the corresponding conformal dimensions are given by:
\begin{equation} \label{eq:h-el}
  h=\bar h=\frac{(2j/n+k)^2}{4p'/p} - \frac{(p-p')^2}{4pp'} \,.
\end{equation}
From the relation~\eqref{eq:ldef} between $n$ and $(p,p')$, it is clear that the electric charge $e=2j/n+k$ spans $\mathbb{Z}/n$ as $j$ ranges from $0$ to $(n-1)$ and $k \in \mathbb{Z}$. Note also that some of these exponents are negative: in particular, the operator $\psi_a^{(0)}$ has electric charge $e=0$ and dimension $h_0=\bar h_0= -\frac{(p-p')^2}{4pp'}<0$, which yields the effective central charge $c_{\rm eff}=c-12(h_0+\bar h_0)=1$. In contrast, the identity operator $\vphi_a^{(\ell)}$ has electric charge $e=2\ell/n$ and dimensions $h=\bar h=0$.

\subsection{Defect operators}

Due to the $n$-periodicity of height variables, the CSOS model also admits some defect operators. For any non-zero integer $m$, the operator that inserts a height defect $a \to a+mn$ around a given point forces $mn$ segments of loops to join in the vicinity of this point. The corresponding `magnetic' conformal dimension is
\begin{equation} \label{eq:h-mag}
  h = \bar h = \frac{p'}{4p}(mn)^2 - \frac{(p-p')^2}{4pp'} \,.
\end{equation}

Electric and magnetic operators may be combined, leading to a (generally non-scalar) primary operator with dimensions
\begin{equation} \label{eq:spectrum}
  \begin{cases}
    h_{em} =  \frac{(ep-mp')^2 - (p-p')^2}{4pp'} \,, \\
    \\
    \bar h_{em} =  \frac{(ep+mp')^2 - (p-p')^2}{4pp'} \,,
  \end{cases}
  \qquad e \in \mathbb{Z}/n \,, \quad m \in n\mathbb{Z} \,.
\end{equation}

\subsection{Partition function on the torus}

Using the approach of~\cite{Pasquier87}, we can map the toroidal partition function of the CSOS model to a linear combination of TL partition functions, which in turn are related to 6V partition functions with proper topological factors~\cite{DSZ87}. This line of arguments yields the simple result in the scaling limit:
\begin{equation}
  Z_{\rm CSOS}(q) = \frac{(q\bar q)^{c/24}}{|\eta(q)|^2} \sum_{e \in \Z/n, \, m \in n\Z} q^{h_{em}} {\bar q}^{\bar h_{em}} \,.
\end{equation}
where $q=e^{2i\pi\tau}$, $\tau$ is the modular parameter, and $\eta$ is the Dedekind eta function. This expression is consistent with the interpretation of the CSOS model as a CFT with central charge~\eqref{eq:cc} and operator content~\eqref{eq:spectrum}.

\end{document}